\documentclass[11pt, a4paper]{article}
\pdfoutput=1
\usepackage{jcappub}
\usepackage{subfig}
\usepackage{caption}
\usepackage{hyperref}
\usepackage{amsmath}
\usepackage{mathtools}
\usepackage{tablefootnote}
\usepackage{soul}
\usepackage{ulem}
\usepackage{color}
\usepackage[usenames,dvipsnames,svgnames,table]{xcolor}
\usepackage{bm}

\definecolor{darkred}{RGB}{135,0,0}

\definecolor{darkgreen}{rgb}{0.0, 0.5, 0.0}

\title{Beware of commonly used approximations I: errors in forecasts}

\author[a,b]{Nicola Bellomo,}
\emailAdd{nicola.bellomo@icc.ub.edu}

\author[c,a,b]{Jos\'e Luis Bernal,}
\emailAdd{jbernal2@jhu.edu}

\author[d,e,f,g,a]{Giulio Scelfo,}
\emailAdd{giulio.scelfo@sissa.it}

\author[h,a]{Alvise Raccanelli,}
\emailAdd{alvise.raccanelli@cern.ch}

\author[a,i]{Licia Verde}
\emailAdd{liciaverde@icc.ub.edu}

\affiliation[a]{ICC, University of Barcelona, IEEC-UB, Mart\'i  i Franqu\`es, 1, E-08028 Barcelona, Spain.}
\affiliation[b]{Dept. de  F\'isica Qu\`antica i Astrof\'isica, Universitat de Barcelona, Mart\'i  i Franqu\`es 1, E-08028 Barcelona, Spain.}  
\affiliation[c]{Department of Physics and Astronomy, Johns Hopkins University, 3400 North Charles Street, Baltimore, Maryland 21218, USA.}
\affiliation[d]{SISSA, Via Bonomea 265, I-34136 Trieste, Italy.}
\affiliation[e]{INFN, Sezione di Trieste, Via Bonomea 265, 34136 Trieste, Italy}
\affiliation[f]{IFPU, Institute for Fundamental Physics of the Universe, via Beirut 2, 34151, Trieste, Italy}
\affiliation[g]{Dipartimento di Fisica e Astronomia G. Galilei, Universit\`a degli Studi di Padova, via F. Marzolo 8, I-35131, Padova, Italy.}
\affiliation[h]{Theoretical Physics Department, CERN, 1 Esplanade des Particules, 1211 Geneva 23, Switzerland.}
\affiliation[i]{ICREA, Pg. Lluis Companys 23, Barcelona, E-08010, Spain.}

\abstract{
In the era of precision cosmology, establishing the correct magnitude of statistical errors in cosmological parameters is of crucial importance. However, widely used approximations in galaxy surveys analyses can lead to parameter uncertainties that are grossly mis-estimated, even in a regime where the theory is well understood (e.g., linear scales). These approximations can be introduced at three different levels: in the form of the likelihood, in the theoretical modelling of the observable and in the numerical computation of the observable. Their consequences are important both in data analysis through e.g., Markov Chain Monte Carlo parameter inference, and when survey instrument and strategy are designed and their constraining power on cosmological parameters is forecasted, for instance using Fisher matrix analyses.
In this work, considering the galaxy angular power spectrum as the target observable, we report one example of approximation for each of such three categories: neglecting off-diagonal terms in the covariance matrix, neglecting cosmic magnification and using the Limber approximation on large scales. We show that these commonly used approximations affect the robustness of the analysis and lead, perhaps counter-intuitively, to unacceptably large mis-estimates of parameters errors (from few~$10\%$ up to few~$100\%$) and correlations. Furthermore, these approximations might even spoil the benefits of the nascent multi-tracer and multi-messenger cosmology. Hence we recommend that the type of analysis presented here should be repeated for every approximation adopted in survey design or data analysis, to quantify how it may affect the results. To this aim, we have developed \texttt{Multi\_CLASS}, a new extension of \texttt{CLASS} that includes the angular power spectrum for multiple (galaxy and other tracers such as gravitational waves) populations. The public release of \texttt{Multi\_CLASS} is associated with this paper.
}

\begin{document}

\begin{flushright}
CERN-TH-2020-044
\end{flushright}

\maketitle

\section{Introduction}
\label{sec:intro}

The next decade promises to be the golden age of galaxy surveys, which unprecedented instrumental sensitivity may enable  potential discoveries of physics beyond the standard, $\Lambda$CDM, cosmological model. This experimental effort will not only improve constraints on the standard cosmological parameters, but also  make possible to explore common extensions of the $\Lambda$CDM, including, for instance, the presence of massive neutrinos or primordial non-Gaussianities.

One of the main challenges the cosmological community will face in the near future is achieving  precision and accuracy at the same time, i.e., extract  the correct value of cosmological parameters from the data, with small, yet accurate, errors. Given the vast amount of data to deal with, sometimes analyses tends to be streamlined, for simplicity and speed, using sets of approximations. However, the robustness of such approximations should be checked (and this has not always been done in the past), since they can introduce systematic errors in the inferred parameters best-fits and errors.

We identify three types  that encompass all commonly taken approximations: on the form of the likelihood used for parameter inference, on the theoretical modelling of a target observable and on its numerical computation. These approximations change the shape of the likelihood in different ways, in particular they can potentially change both the position of the maximum and the curvature of the likelihood around it. The first effect induces a shift in the inferred best-fit parameters, and it is investigated in a companion paper~\cite{bernal:bestfitshift}. The second one changes the magnitude of the inferred errors and the direction of degeneracies in parameter space, hence it introduces \textit{errors on the inferred errors}.

In this work we focus on the effects of mis-estimating the likelihood curvature around its maximum. These effects appear both in the exploration of the parameter posterior distribution, as done in a typical MCMC analysis, and when forecasting the constraining power of a future survey, typically using a Fisher matrix approach~\cite{fisher:fisher, bunn:fisher, vogeley:fisher, tegmark:fisher, heavens:fishermatriximprovement, coe:fishermatrixguide}. This second aspect is the subject of our study here, given that the estimation of the maximum constraining power of an experiment is a fundamental process in the design of the experiment itself, the instrument and the survey strategy. Nevertheless we stress that the same effect would happen also in data analysis if the same approximations are used.

In the spirit of using quick and easy forecasting tools, the modelling of the likelihood and  the observable is usually simplified to reduce mathematical complexity and computational time. First, in the case of the likelihood modelling, the covariance between different data sets is usually neglected, estimating poorly the correlation existing between target observables. This approximation obviously fails for instance when different tracers (e.g., different galaxy populations) measured by the same experiment are used to measure the same observable in the same volume of the Universe (the so called {\it multi-tracer} approach). Moreover, it fails also when one tracer is measured by two different experiments in the same patch of the sky. Second, certain physical effects are neglected or overlooked if they do not directly depend on the cosmological parameters of interest. Third, some numerical approximations are used in regimes where they break down.

The fact that the Fisher matrix analysis has some limitations~\cite{joachimi:fishermatriximprovement, sellentin:nglkl, sellentin:fishermatriximprovement, amendola:fishermatriximprovement} is used to justify the use of inaccurate modelling. In fact, it is typically assumed that these approximations will not significantly affect the forecasted error-bars, even though they might bias the result of real data analysis. However, it is challenging to estimate a priori if the error introduced is comparable to the intrinsic error of a Fisher forecast. Moreover, inaccurate forecasts may lead to a wrong estimation of parameters covariance matrices, changing the final estimated error and correlations between parameters. Since experiments are designed to achieve a target sensitivity and to break existing parameter degeneracies, mis-estimating the errors of cosmological parameters and degeneracies between them might  hinder the entire science-case for the experiment. 

Given the potential of current and forthcoming galaxy surveys such as EMU~\cite{norris:emu}, DESI~\cite{aghamousa:desi}, Euclid~\cite{laureijs:euclidsciencebook}, LSST~\cite{abell:lsstsciencebook}, SPHEREx~\cite{dore:spherexwhitepaperI}, WFIRST~\cite{spergel:wfirstwhitepaper} and SKA~\cite{maartens:ska, bacon:skaredbook}, here we focus on galaxy clustering at large scales. We consider the galaxy angular power spectrum as target observable. First we show the potential problems arising from using a wrong covariance matrix in the likelihood. Then we analyse one of the most often neglected physical effects, i.e., cosmic magnification, and one of the most common numerical approximations, i.e., the Limber approximation, which breaks down at large scales. 

Finally, we go beyond the ``traditional'' single-tracer analysis and we consider the combination of different tracers. Different galaxy surveys trace different galaxy types which, in turn, trace the underlying density field in slightly different ways. The technique of combining different tracers has the potential to reduce cosmic variance~\cite{seljak:cosmicvariance, mcdonald:cosmicvariance, blake:gamamultitracer, beutler:bossmultitracerI, marin:bossmultitracerII}, hence its great importance for cosmology. While the multi-tracer approach can enhance the amount of information from a given survey, the power of this method and the robustness of its results might be spoiled by the very same kind of approximations mentioned above. 

We developed a new extension of \texttt{CLASS}~\cite{blas:class, didio:classgal} called \texttt{Multi\_CLASS}\footnote{The code will be publicly released after the article is accepted. Users can find and download the code in the GitHub page \url{https://github.com/nbellomo/Multi_CLASS}.}, to include the multi-tracer case in our analysis. \texttt{Multi\_CLASS} is the first public code that allows to compute the angular power spectrum for multiple galaxy (and other tracers including gravitational waves events) populations. The code allows the user to specify, for each tracer, its number density redshift distribution, bias, magnification bias and evolution bias. Moreover, we implemented also the effect of primordial non-Gaussianity of the local-type, parametrised by~$f_\mathrm{NL}$, on the tracer bias.

The paper does not aim to provide for specific experiments a quantitative estimate of the errors induced by a wrong modelling, since this is strongly case-dependent. The actual purpose of this work is to show that forecasts are unreliable if important physical effects are neglected (even when they do not depend on the cosmological parameter of interest and thus this may appear counter intuitive) or when the approximations adopted are not sufficiently accurate in the regime under study.

The paper is structured as follows: in section~\ref{sec:observable_modeling_fisher_matrix} we introduce the galaxy angular power spectrum (i.e., our \textit{observable}) and the Fisher matrix formalism, while in section~\ref{sec:tools_cosmomodel} we introduce a set of diagnostic tools and the experimental set-up. We show the effects of an inaccurate modelling of the likelihood and of the observable in sections~\ref{sec:likelihood_modeling} and~\ref{sec:signal_modeling}, respectively. In section~\ref{sec:multitracing} we present \texttt{Multi\_CLASS} and we study the impact of approximations when multiple tracers are considered. Finally we conclude in section \ref{sec:conclusions}. Appendix~\ref{app:relativistic_number_counts} contains details on the galaxy number count power spectrum, while appendix~\ref{app:multi_class} is dedicated to describe \texttt{Multi\_CLASS}, both for users and for potential developers.

%%%%%%%%%%%%%%%%%%%%%%%%%%%%%%%%%%%%%%%%%%%%%%%%%%%%%%%%%%%%%%%%%%%

\section{Galaxy angular power spectrum and likelihood modelling}
\label{sec:observable_modeling_fisher_matrix}

While this section is mostly of review, it serves to define all the quantities used.  We introduce the modelling of galaxy clustering in harmonic space in \S~\ref{subsec:galaxy_clustering_theory}, and the relevant likelihood and Fisher matrix formalism in \S~\ref{subsec:likelihood_fisher_matrix}. The impatient reader can skim this section and then go directly to section~\ref{sec:tools_cosmomodel}. For the reader planning to use \texttt{Multi\_CLASS}, this section is key to walk through the structure of the code and the implementation of relevant equations in it.

%%%%%%%%%%%%%%%%%%%%%%%%%%%%%%%%%%%%%%%%%%%%%%%%%%%%%%%%%%%%%%%%%%%

\subsection{Theoretical modelling of galaxy clustering}
\label{subsec:galaxy_clustering_theory}

Testing cosmological models using galaxy clustering is one of the main goals of current and future astrophysical experiments. Galaxy clustering can be studied using a variety of statistical methods, usually focused on the two- and three-point correlations in configuration, Fourier or harmonic spaces. In this work we consider the galaxy angular power spectrum as our observable, i.e., the two-point statistics of the observed galaxy number count fluctuation in harmonic space. One of the advantages of this methodology is the possibility to easily account for spherical symmetry of the sky at large scales (i.e., going beyond the flat-sky, distant observer approximations, which do not hold for future surveys covering a large fraction of the celestial sphere). As new surveys observe larger and larger fractions of the sky, it becomes mandatory to drop the widely used distant observer/flat sky approximation. Methods to do so in configuration and spherical harmonics spaces have been developed, see e.g., refs.~\cite{heavens:rsdspherical, szalay:wideangle, matsubara:wideangle, papai:wideangle, raccanelli:wideangle, bertacca:wideangle, raccanelli:wideangleII, dai:tamwaves, yoo:wideangle, blake:wideangle, taruya:wideangle}.

We can expand the galaxy number count fluctuation in spherical harmonics as
\begin{equation}
\delta^X(z,\hat{\mathbf{n}}) = \sum_{\ell m} a^{X,z}_{\ell m} Y_{\ell m}(\hat{\mathbf{n}}),
\label{eq:harmonic_expansion}
\end{equation}
where $a^{X,z}_{\ell m}$ are the spherical harmonics coefficients of the tracer $X$ at redshift $z$ along the line of sight $\hat{\mathbf{n}}$, and $Y_{\ell m}$ are spherical harmonics. The angular power spectrum~$C^{XY}_\ell(z_i,z_j)$ of tracers~$X$ and~$Y$ at redshift~$z_i$ and~$z_j$, respectively, is computed from the spherical harmonics coefficients as~\cite{raccanelli:crosscorrelation, pullen:crosscorrelation}
\begin{equation}
\left\langle a^{X,z_i}_{\ell m}a^{Y,z_j*}_{\ell'm'} \right\rangle = \delta^K_{\ell\ell'} \delta^K_{mm'} C^{XY}_\ell(z_i,z_j),
\label{eq:angular_power_spectrum}
\end{equation}
where $\delta^K$ stands for the Kronecker delta and $^*$ denotes the complex conjugate. Galaxies are discrete objects, hence the observed galaxy power spectrum has to include also a shot noise term. In the following we consider only a scale-independent shot noise contribution to the theoretical galaxy angular power spectrum. Therefore we define the total angular power spectrum as
\begin{equation}
\tilde{C}^{XY}_\ell(z_i,z_j) = C^{XY}_\ell(z_i,z_j) + \frac{\delta^K_{ij}\delta^K_{XY}}{dN_X(z_i)/d\Omega},
\label{eq:total_angular_power_spectrum}
\end{equation}
where $dN_X(z_i)/d\Omega$ is the average tracer number density per steradian. Notice that for different tracers $(X\neq Y)$ or different redshifts $(i\neq j)$ there is no shot noise. Shot noise is not the only possible source of noise: the total noise can also depend on the specifics of the instrument and on our theoretical understanding of the observable. These effects can be added as extra terms to equation~\eqref{eq:total_angular_power_spectrum}. For our purposes it is sufficient to consider only shot noise, whereas in real data analysis these extra contributions must be taken into account.

Following the notation of ref.~\cite{bonvin:cl}, the angular power spectrum can be written as
\begin{equation}
C^{XY}_\ell(z_i,z_j) = 4\pi\int\frac{dk}{k} \mathcal{P}(k) \Delta^{X,z_i}_{\ell}(k) \Delta^{Y,z_j}_{\ell}(k),
\label{eq:Cls}
\end{equation}
where $\mathcal{P}(k) = k^3P(k)/2\pi^2$ is the almost scale-invariant primordial power spectrum and the latter two terms of the integrand read as
\begin{equation}
\Delta^{X,z_i}_{\ell}(k) = \int_{0}^{\infty} dz \frac{dN_X}{dz}W(z,z_i,\Delta z_i)\Delta^X_\ell(k,z),
\label{eq:Delta_l}
\end{equation}
where we have introduced a window function~$W(z,z_i,\Delta z_i)$, centred at redshift~$z_i$ with half-width~$\Delta z_i$,\footnote{The width of the window function implicitly defines the width of the redshift bin, hence for practical purposes speaking of bin width or window function width is equivalent. For instance in the case of a top-hat window function we have~$W(z,z_j,\Delta z_j)\propto \Theta\left(z-(z_j-\Delta z_j)\right)\Theta\left(z_j+\Delta z_j-z)\right)$, where $\Theta$ is the Heaviside function, while for a Gaussian window function we have~$W(z,z_j,\Delta z_j)\propto \exp \left[-\frac{1}{2}\left(\frac{z-z_j}{\Delta z_j}\right)^2 \right]$.} the tracer number density per redshift interval~$dN_X/dz$ and the number count fluctuation transfer function~$\Delta^X_\ell(k,z)$. The integral of~$W(z,z_i,\Delta z_i) dN_X/dz$ is normalized to unity. In general, the total observed number count fluctuation receives contributions from density ($\mathrm{den}$), velocity ($\mathrm{vel}$), lensing ($\mathrm{len}$) and gravity ($\mathrm{gr}$) effects and reads as~\cite{bonvin:cl, challinor:deltag}
\begin{equation}
\Delta^X_\ell (k, z) = \Delta^{X,\mathrm{den}}_{\ell}(k, z) + \Delta^{X,\mathrm{vel}}_\ell(k, z) + \Delta^{X,\mathrm{len}}_\ell(k, z) + \Delta^{X,\mathrm{gr}}_\ell(k, z),
\label{eq:numbercount_fluctuation}
\end{equation}
where the explicit form of all the above contributions is reported in appendix~\ref{app:relativistic_number_counts}. The nomenclature reflects straightforwardly how these contributions are implemented in \texttt{CLASS}. Every tracer $X$ is characterized by a set of parameters, namely a total bias parameter~$b_{X,\mathrm{tot}}$, a magnification bias parameter~$s_X$ and an evolution bias parameter~$f^\mathrm{evo}_X$, all of them possibly scale- and redshift-dependent. These parameters enter in the RHS of equation~\eqref{eq:numbercount_fluctuation} and regulate the amplitude of the different contributions (see also appendix~\ref{app:relativistic_number_counts}). In the following we briefly describe the physical origin of these parameters.

The relation between observable tracers of large-scale structure and the underlying matter distribution is called the large-scale structure biasing (see e.g., refs.~\cite{mo:dmhaloclustering, matarrese:clusteringevolution, dekel:stochasticbiasing, benson:galaxybias, peacock:halooccupation}, and ref.~\cite{desjacques:biasreview} for a review). Despite tracing the same underlying matter distribution, different tracers can have different clustering properties. We quantify these different clustering properties through the total galaxy bias parameter~$b_{X,\mathrm{tot}}$, which governs the ratio of clustering amplitude of the selected tracer to that of the dark matter.

The modelling of galaxy bias have been the subject of extensive studies in the last years~\cite{mcdonald:dmclustering, assassi:halobias, senatore:galaxybias, mirbabayi:galaxybias, saito:nonlocalhalobias, desjacques:galaxybias, schmittfull:galaxybias, fujita:galaxybias}, however a complete theory of galaxy biasing has not been fully developed yet. For the scope of this work we adopt a simplified, widely used, model: we consider only a first-order large-scale linear bias, neglecting higher order contributions and effects due to the presence of e.g., massive neutrinos and other relics~\cite{raccanelli:neutrinobias, munoz:neutrinobias, vagnozzi:neutrinobias, valcin:behappy}. Thus, we assume a redshift-dependent total bias with scale dependence given only by primordial non-Gaussianity contributions as in~\cite{matarrese:highredshiftobjects, dalal:ngbias, matarrese:nongaussianbias}
\begin{equation}
b_{X,\mathrm{tot}}(k,z) = b_X + 2(b_X-1)f_\mathrm{NL}\delta_\mathrm{crit}\frac{3\Omega_{m0}H_0^2}{2c^2k^2T(k)D(z)},
\label{eq:total_galaxy_bias}
\end{equation}
where $b_X$ is the Eulerian Gaussian galaxy bias (here for illustrative purposes we assume it to be scale- and redshift-independent), $f_\mathrm{NL}$ is the amplitude of primordial non-Gaussianities of the local type,\footnote{In equation~\eqref{eq:total_galaxy_bias} and in the rest of the paper we use the LSS convention for the amplitude of primordial non-Gaussianities, i.e., $f_\mathrm{NL}\equiv f^\mathrm{LSS}_\mathrm{NL}$. However, for the sake of conciseness, we omit the $^\mathrm{LSS}$ superscript.} $\delta_\mathrm{crit}$ is the critical threshold associated to gravitational collapse ($\delta_\mathrm{crit}=1.686$, assuming spherical collapse in an Einstein-de Sitter cosmology), $\Omega_{m0}$ is the present-day matter fractional density, $H_0$ is the present-day Hubble expansion rate, $c$ is the speed of light, $T(k)$ is the matter transfer function and $D(z)$ is the linear growth factor normalized to unity at redshift~$z=0$.\footnote{Equation~\eqref{eq:total_galaxy_bias} is derived assuming an Einstein-de Sitter cosmology, in which we can separate the scale evolution, represented by~$T(k)$, from the time evolution, described by~$D(z)$. In different scenarios, as the one in which neutrinos are massive, this separation would not be possible and a scale- and redshift-dependent ``transfer function''~$\mathcal{T}(k,z)$ should be used. Moreover, comparison with N-body simulations~\cite{grossi:pngnbody, wagner:pngnbody} indicate that there might be a correction factor of order unity to equation~\eqref{eq:total_galaxy_bias}. Therefore $f_\mathrm{NL}$ should be interpreted as an effective parameter of about the same order of magnitude of the true~$f_\mathrm{NL}$.}

Magnification lensing changes the source number count surface density on the sky in two competing ways~\cite{turner:magnificationbias}: by increasing the area, which in turn decreases the projected number density, but also by magnifying individual sources and promoting faint objects above the survey magnitude limit. These effects change the observed number density $n_\mathrm{obs}$ in a flux-limited survey. At linear order (see e.g., ref.~\cite{bohm:lensingcorrections} for higher-order corrections), this correction reads as
\begin{equation}
n_\mathrm{obs} = n_X \left[ 1 + (5s_X-2) \kappa \right],
\end{equation}
where $n_X$ is the intrinsic tracer number density, $s_X$ is called magnification bias parameter and $\kappa = \frac{1}{2}\nabla^2\psi$ is the convergence~\cite{bartelmann:convergence}, namely an isotropic change of the source size generated by the lensing potential~$\psi$. If the tracers of interest are galaxies, then the change in the number of observed sources depends on the value of the slope of the faint-end of the luminosity function~\cite{matsubara:lensing, hui:magnificationbias, liu:magnificationbias, montanari:magnificationbias}
\begin{equation}
s_X(z) = \left.\frac{d\log_{10}\frac{d^2N_X(z,m<m_\mathrm{lim})}{dzd\Omega}}{dm}\right|_{m_\mathrm{lim}},
\label{eq:galaxymagnificationbias}
\end{equation}
where $m$ is the apparent magnitude, $m_\mathrm{lim}$ is the magnitude limit of the survey and $d^2N_X/dzd\Omega$ is the tracer number density per redshift interval per steradian. Following the same logic, the definition above can be adapted also for other tracers, such as gravitational waves, as done in ref.~\cite{scelfo:gwxlss}. In both cases the magnification bias parameter mainly contributes to the lensing part, although it is also present in the velocity and gravity terms. The reader should keep in mind that the specific value $s_X=0.4$ is associated to a compensation between the two competing effects. The lensing contribution vanishes for this value of the magnification bias parameter.

Furthermore the number of tracers is not necessarily conserved as function of redshift, e.g., galaxies can form, therefore the tracer number density might not scale as $a^{-3}$ with the scale factor $a$. To account for the evolution of the number distribution of tracers, we include the so-called evolution bias $f^\mathrm{evo}_X$ defined as \cite{challinor:deltag, jeong:evolutionbias, bertacca:wideangle}
\begin{equation}
f^\mathrm{evo}_X(z) = \frac{d\log\left(a^3\frac{d^2N_X}{dzd\Omega}\right)}{d\log a}.
\label{eq:evolution_bias}
\end{equation}
This term enters in the velocity and gravity contributions in equation~\eqref{eq:numbercount_fluctuation}. We use the observed number density instead of the true one in the definition of evolution bias. For our purposes this is adequate because this parameter enters in subleading terms of equation~\eqref{eq:numbercount_fluctuation} and the uncertainties in the modelling of the evolution of the tracers are significant.

%%%%%%%%%%%%%%%%%%%%%%%%%%%%%%%%%%%%%%%%%%%%%%%%%%%%%%%%%%%%%%%%%%%

\subsection{Likelihood and Fisher matrix}
\label{subsec:likelihood_fisher_matrix}

Consider a vector~$\mathbf{D}$ containing the data, a mean vector~$\bm{\mu}=\left\langle\mathbf{D}\right\rangle$ and a covariance matrix~$\mathrm{Cov} = \left\langle(\mathbf{D}-\bm{\mu})(\mathbf{D}-\bm{\mu})^\dagger\right\rangle$, where angle brackets $\langle\ \cdot\ \rangle$ indicate the statistical expectation value and the dagger~$^\dagger$ indicates the complex conjugate plus transpose operation. A Gaussian likelihood $\mathcal{L}$ can be written in full generality as
\begin{equation}
-2\log\mathcal{L} \propto \log\det(\mathrm{Cov}) + (\mathbf{D}-\bm{\mu})^\dagger \mathrm{Cov}^{-1} (\mathbf{D}-\bm{\mu}),
\label{eq:gaussian_likelihood}
\end{equation}
where typically both the mean vector and the covariance depend on the set of model parameters~$\left\lbrace\theta_\alpha\right\rbrace$, and we choose to omit the constant factors appearing in the likelihood. 

When computing two-point statistics in harmonic space, one can choose to work either with the spherical harmonics coefficients or the angular power spectra as data, even if strictly speaking only the former are real Gaussian variables. In the following we assume, for simplicity, full sky coverage, to avoid introducing correlation between different multipoles. The data can be organised either in a column vector~$\bm{a}^\mathrm{data}_{\ell m}$ or in a column vector~$\bm{C}^\mathrm{data}_\ell$, depending on whether we work with spherical harmonics coefficients or angular power spectra, respectively. The vectors~$\bm{a}^\mathrm{data}_{\ell m}$ and~$\bm{C}^\mathrm{data}_\ell$ are not independent if they are relative to the same field:
\begin{equation}
\bm{a}^\mathrm{data}_{\ell m} = 
\left( \begin{matrix}
a^{(1)}_{\ell m} \\
a^{(2)}_{\ell m} \\
\vdots	\\
a^{(N)}_{\ell m}
\end{matrix} \right),
\quad
\bm{C}^\mathrm{data}_\ell = \frac{1}{2\ell+1}
\left( \begin{matrix}
\sum_{m=-\ell}^{+\ell}a^{(1)}_{\ell m}a^{(1)*}_{\ell m} \\
\sum_{m=-\ell}^{+\ell}a^{(1)}_{\ell m}a^{(2)*}_{\ell m} \\
\vdots	\\
\sum_{m=-\ell}^{+\ell}a^{(N)}_{\ell m}a^{(N)*}_{\ell m}
\end{matrix} \right),
\label{eq:vectors}
\end{equation}
where the index~$j\in\{1,2,\cdots,N\}$ of the partial wave coefficients~$a^{(j)}_{\ell m}$ refer to different redshift bins and/or different tracers, depending on the specific case. The two vectors of equation~\eqref{eq:vectors} have different dimensions: here we assume~$N=\dim(\bm{a}^\mathrm{data}_{\ell m})$ independent partial wave coefficients (for each $\ell m$ pair) that generate~$N(N+1)/2=\dim(\bm{C}^\mathrm{data}_\ell)$ independent angular power spectra (for each $\ell$).

The above description is valid for any number of tracers and/or redshift bins. In the standard case of a single tracer~$X$ in~$N_X$ redshift bins, we have~$N=N_X$. On the other hand, for two tracers~$X$ and~$Y$ divided in~$N_X$ and~$N_Y$ redshift bins, we have~$N=N_X+N_Y$. In this case the~$\bm{C}^\mathrm{data}_\ell$ data vector consists of~$N_X(N_X+1)/2+N_Y(N_Y+1)/2$ auto-tracer angular power spectra ($C^{XX}_\ell$ and~$C^{YY}_\ell$) and by~$N_X\times N_Y$ cross-tracer angular power spectra ($C^{XY}_\ell$). In this work we consider different populations of galaxies as tracers, however the framework described in this section can also incorporate tracers which are not galaxies, as in the case of cross-correlating large-scale structure with the cosmic microwave background~\cite{ho:cmbxlss, hirata:cmbxlss, raccanelli:crosscorrelation}, with gravitational waves~\cite{camera:gwlensing, scelfo:gwxlss, bertacca:asgwb, calore:gwxlss, alonso:gwxlss}, with neutrinos~\cite{fang:neutrinosxlss}, with ultra-high energy cosmic rays~\cite{urban:uhecrxlss, motloch:uhecrxlss} and so on.

If we choose to work with the spherical harmonics coefficients as data, we have $\mathbf{D}=\bm{a}^\mathrm{data}_{\ell m}$, $\bm{\mu}=0$ and $\mathrm{Cov}\equiv\mathcal{C}_\ell(\{\theta\})$, therefore equation~\eqref{eq:gaussian_likelihood} reads as
\begin{equation}
-2\log\mathcal{L} \propto \sum_\ell\sum_{m=-\ell}^{+\ell} \left[\log\det \mathcal{C}_\ell(\{\theta\}) + \left(\bm{a}^\mathrm{data}_{\ell m}\right)^\dagger \mathcal{C}^{-1}_\ell(\{\theta\}) \bm{a}^\mathrm{data}_{\ell m}\right],
\label{eq:gaussian_likelihood_Deltalm}
\end{equation}
where in this case only the elements of the covariance matrix $\left(\mathcal{C}_\ell\right)_{IJ} = \tilde{C}^{IJ}_\ell$ depend on the cosmological parameters. On the other hand, considering the angular power spectra as data vector, equation~\eqref{eq:gaussian_likelihood} reads as
\begin{equation}
-2\log\mathcal{L} \propto \sum_\ell\left[\log\det \mathcal{M}_\ell(\{\theta\}) + \left(\bm{C}^\mathrm{data}_\ell-\bm{C}_\ell(\{\theta\})\right)^T \mathcal{M}^{-1}_{\ell}(\{\theta\}) \left(\bm{C}^\mathrm{data}_\ell-\bm{C}_\ell(\{\theta\})\right)\right],
\label{eq:gaussian_likelihood_Cl}
\end{equation}
where $^T$ denotes the transpose operator, $\mathbf{D}=\bm{C}^\mathrm{data}_\ell$ and both the mean $\bm{\mu}=\bm{C}_\ell(\{\theta\})\neq \mathbf{0}$ and the covariance matrix $\mathrm{Cov}\equiv \mathcal{M}_\ell(\{\theta\})$ depend on the cosmological parameters. The elements of the covariance matrix can be calculated from the definition given above equation~\eqref{eq:gaussian_likelihood} for the general case using the Wick theorem. We can associate to every index $I$ and $J$ of the column vector $\bm{C}_\ell$ a couple of indexes $(I_1,I_2)$ and $(J_1,J_2)$, representing the two indexes of the spherical harmonics coefficients appearing in equation~\eqref{eq:vectors} that generate such angular power spectra, i.e., $(\bm{C}_\ell)_I = C^{(I_1,I_2)}_\ell = \sum_m a^{(I_1)}_{\ell m}a^{(I_2)*}_{\ell m} / (2\ell + 1)$. Hence the element $IJ$ of the covariance matrix reads as
\begin{equation}
\left(\mathcal{M}_{\ell}\right)_{IJ} = \frac{1}{2\ell+1}\left[\tilde{C}^{(I_1,J_1)}_\ell\tilde{C}^{(I_2,J_2)}_\ell+\tilde{C}^{(I_1,J_2)}_\ell\tilde{C}^{(I_2,J_1)}_\ell\right],
\label{eq:Mell_covariance_matrix}
\end{equation}
where we used the total angular power spectra defined in equation~\eqref{eq:total_angular_power_spectrum}. 

Despite the appearances, the two likelihoods in equations~\eqref{eq:gaussian_likelihood_Deltalm} and~\eqref{eq:gaussian_likelihood_Cl} are not equivalent: they do not contain the same amount of information. In fact it can be shown that if the $a_{\ell m}$ are Gaussian random variables, then the angular power spectra follow a Wishart distribution~\cite{wishart:distribution}, which can be approximated as a Gaussian distribution only at high multipoles (i.e., small scales), where the central limit theorem applies. A Wishart distribution has non-zero skewness: even though the expectation value of the angular power spectra data vector is $\left\langle \mathbf{C}^\mathrm{data}_\ell\right\rangle = \mathbf{C}_\ell(\{\theta\})$ for each $\ell$, the maximum of the likelihood is located at $\left(1-\ell^{-1}\right)C_\ell(\{\theta\})$. Therefore working with a Gaussian likelihood at low multipoles, or equivalently at large scales, biases the final answer, whereas at high multipole the $\ell^{-1}$ correction is negligible. Several approximations have been proposed  to correct for this effect, see e.g., refs.~\cite{bond:nglklI, bond:nglklII, verde:nglkl, smith:nglkl, percival:nglkl}.

The standard Fisher analysis assumes Gaussian likelihood and errors. The approach relies on the estimation of the log-likelihood curvature around its maximum and it returns the smallest error we can hope to achieve, known also as the Cram\'er-Rao bound. The curvature of the likelihood around the maximum is estimated using a Taylor expansion up to second order, while higher order terms are typically neglected. The procedure is clearly idealized in terms of knowledge of the parameter posterior (which is not always Gaussian) and in terms of characterization of the instrument and eventual observational systematics, even if this kind of analysis can be used to quantify the effects of nuisance parameters. Therefore it rarely reflects the performance of real experiments and the actual errors are usually larger than forecasted estimates, despite several improvements that have been made, for instance to take into account non-Gaussian posteriors, see e.g., refs.~\cite{joachimi:fishermatriximprovement, sellentin:nglkl, sellentin:fishermatriximprovement, amendola:fishermatriximprovement}. The method still remains an useful and easy-to-implement technique to compare performances of different instruments or different survey strategies for a given instrument. Moreover, confidence regions derived from the Fisher matrix usually provide a reasonable estimate of parameters errors and degeneracies. 

Elements of the Fisher matrix are obtained as the second derivative of the log-likelihood with respect to the parameters of the model. The Fisher element corresponding to the~$\theta_\alpha$ and~$\theta_\beta$ parameters is 
\begin{equation}
F_{\alpha\beta} = \left\langle - \frac{\partial^2\log \mathcal{L}}{\partial \theta_{\alpha}\partial \theta_\beta}\right\rangle.
\label{eq:fisher_matrix_definition}
\end{equation}
The parameter covariance matrix~$\Sigma$ is the inverse of the Fisher matrix, i.e., $\Sigma=F^{-1}$. From the Fisher matrix we can extract two types of errors on parameters: conditional errors and marginal errors. The conditional error on~$\theta_\alpha$ is given by $\sigma^\mathrm{cond.}_{\theta_\alpha}=1/\sqrt{F_{\alpha\alpha}}$ and it represents the error obtained keeping all the parameters fixed except~$\theta_\alpha$. On the other hand, the marginal error on $\theta_\alpha$ is  $\sigma^\mathrm{marg.}_{\theta_\alpha}=\sqrt{(F^{-1})_{\alpha\alpha}}$ and it is the error obtained when estimating all the parameters simultaneously (i.e., marginalised over all other parameters). In the rest of the work we focus only on marginal errors. Note that bigger Fisher matrix elements $F_{\alpha\beta}$ are broadly associated to smaller error for the parameters. However, this association is not so straightforward for marginal errors for non-diagonal matrices because of the matrix inversion operation.

The Fisher matrix obtained from the likelihood in equation~\eqref{eq:gaussian_likelihood_Deltalm} is 
\begin{equation}
F_{\alpha\beta} = \sum_\ell  \frac{2\ell+1}{2} \mathrm{Tr}\left[\frac{\partial \mathcal{C}_\ell}{\partial\theta_\alpha}\mathcal{C}^{-1}_\ell\frac{\partial \mathcal{C}_\ell}{\partial\theta_\beta}\mathcal{C}^{-1}_\ell\right] = \sum_\ell \frac{\partial \bm{C}^T_\ell}{\partial\theta_\alpha} \mathcal{M}^{-1}_\ell \frac{\partial \bm{C}_\ell}{\partial\theta_\beta},
\label{eq:fisher_likelihood_delta}
\end{equation}
where $\mathrm{Tr}[\ \cdot\ ]$ indicates the trace operator and we have used matrices properties to write two equivalent forms of the Fisher matrix commonly found in literature (see appendix~A of ref.~\cite{hamimeche:cmblikelihood}). On the other hand, the Fisher matrix for the likelihood in equation~\eqref{eq:gaussian_likelihood_Cl} is given by
\begin{equation}
\begin{aligned}
F_{\alpha\beta} &= \sum_\ell \frac{1}{2} \mathrm{Tr}\left[\frac{\partial \mathcal{M}_\ell}{\partial\theta_\alpha}\mathcal{M}^{-1}_\ell\frac{\partial \mathcal{M}_\ell}{\partial\theta_\beta}\mathcal{M}^{-1}_\ell + \mathcal{M}^{-1}_\ell \left( \frac{\partial \bm{C}_\ell}{\partial\theta_\alpha}\frac{\partial \bm{C}^T_\ell}{\partial\theta_\beta} + \frac{\partial \bm{C}_\ell}{\partial\theta_\beta}\frac{\partial \bm{C}^T_\ell}{\partial\theta_\alpha} \right) \right]	\\
&= \sum_\ell \frac{1}{2} \mathrm{Tr}\left[\frac{\partial \mathcal{M}_\ell}{\partial\theta_\alpha}\mathcal{M}^{-1}_\ell\frac{\partial \mathcal{M}_\ell}{\partial\theta_\beta}\mathcal{M}^{-1}_\ell \right] + \sum_\ell \frac{\partial \bm{C}^T_\ell}{\partial\theta_\alpha} \mathcal{M}^{-1}_\ell \frac{\partial \bm{C}_\ell}{\partial\theta_\beta}.
\end{aligned}
\label{eq:fisher_likelihood_cl}
\end{equation}
Equation~\eqref{eq:fisher_likelihood_cl} contains an extra term with respect to the Fisher matrix of equation~\eqref{eq:fisher_likelihood_delta}. This extra term comes from the fact that the covariance matrix $M_\ell$ depends on cosmological parameters. As noticed  by ref.~\cite{carron:gaussianassumption}, that term leads to an overestimate of the amount of information contained in the data. Hence, it induces a violation of the Cram\'er-Rao bound and it overestimates the real constraining power of a survey, which is given by the second term of the RHS of equation~\eqref{eq:fisher_likelihood_cl} or by the RHS of equation~\eqref{eq:fisher_likelihood_delta}. Therefore, it is more correct to ignore the dependence of the covariance matrix on cosmological parameters when using a Gaussian likelihood for the angular power spectra. This extra term contributes more significantly at low multipoles; once high multipoles contributions are included, the choice of including or not a parameter dependence in the covariance matrix becomes irrelevant, as noticed in ref.~\cite{kodwani:parametersincovariance}. Note also that the effects of a parameter-dependent covariance matrix have already been analysed in the context of cosmic shear \cite{eifler:parametersincovariance} and baryon acoustic oscillations~\cite{labatie:parametersincovariance}, finding that the effects can be accounted for by suitably rescaling the contours by a numerical factor.

In the computation of the Fisher matrix, either in equation~\eqref{eq:fisher_likelihood_delta} or~\eqref{eq:fisher_likelihood_cl}, an additional parameter~$f_\mathrm{sky}$ is typically introduced to account for a fractional coverage of the sky and/or the effect of the mask on scales much smaller than the ones of the mask. Since partial sky coverage induces mode-coupling between different multipoles, the two likelihoods would need a different form to account for this effect. For this reason,  here we choose to work assuming full-sky coverage, ~$f_\mathrm{sky}=1$.

Fisher matrices are useful also to compare the constraining power of different experiments on specific parameters; using  marginalised forecast uncertainties it is customary to compute the so called Figure-of-Merit (FoM). Figures-of-Merit are usually defined for a given pair of parameters as the reciprocal of the area of the error ellipse enclosing the $95\%$ confidence limit in the two parameters plane~\cite{albrecht:figureofmeritI, wang:figureofmerit, albrecht:figureofmeritII} marginalised over all other parameters. Broadly speaking, larger FoM indicates greater accuracy. Moreover, within the Fisher matrix formalism, it is also possible to quantify the shift in parameter estimate caused by a wrong assumption on a set of fiducial parameters~\cite{kim:errorsonparameters, taylor:errorsonparameters, heavens:errorsonparameters} or due to an inaccurate modelling of the observed signal~\cite{knox:parametershift, taruya:parametershift, duncan:parametershift, natarajan:parametershift, pullen:parametershift, cardona:parametershift}, which in turn affects also Bayesian model selection. We refer the interested reader to ref.~\cite{bernal:bestfitshift}, where these aspects are described and analyses in details, and where a generalised approach to all these issues is developed.

%%%%%%%%%%%%%%%%%%%%%%%%%%%%%%%%%%%%%%%%%%%%%%%%%%%%%%%%%%%%%%%%%%%

\subsection{Analyses of sky maps with different multipole ranges}
\label{subsec:sky_maps_analysis}

The model presented in \S~\ref{subsec:galaxy_clustering_theory} is accurate only in the linear regime, hence it should be used to study galaxy clustering only at linear scales. Therefore we need to identify in which range of multipoles we can safely assume that non-linearities play a marginal role, i.e., up to which maximum multipole~$\ell_\mathrm{max}$ we trust our theoretical model.

Suppose for simplicity to have a single-tracer galaxy survey and to have a set of~$N$ sky maps of galaxy distribution at different mean redshift $\{z_1, ..., z_N\}$. We define for every sky map, i.e., for redshift bin, a maximum multipole $\ell_{\mathrm{max},j} \equiv \ell_\mathrm{max}(z_j)$. Thus, as explained in equation~\eqref{eq:harmonic_expansion}, we expand in spherical harmonics the galaxy number count fluctuation of every sky map as
\begin{equation}
\delta(z_j,\hat{\mathbf{n}}) = \sum_{\ell=\ell_\mathrm{min}}^{\ell_{\mathrm{max},j}} \sum_{m=-\ell}^{+\ell} a^{z_j}_{\ell m} Y_{\ell m}(\hat{\mathbf{n}}),
\end{equation}
for $j=1, 2, ..., N$. 

Suppose now one wants to write a Gaussian likelihood for the spherical harmonics coefficients. In this case equation~\eqref{eq:gaussian_likelihood_Deltalm} cannot be used since, for instance, $a^{z_1}_{\ell m} \equiv 0$ for all the multipoles $\ell_{\mathrm{max},1} < \ell \leq \ell_{\mathrm{max},N}$: the covariance matrix, having one row and one column of zeros, would be singular. However, it is still possible to build a Gaussian likelihood using maps with equal multipole ranges. The total likelihood for the set of~$N$ sky maps is given by the product of $N$ different likelihoods, i.e., $\mathcal{L}_\mathrm{tot} = \mathcal{L}_1 \times \mathcal{L}_2 \times \cdots \times \mathcal{L}_N$, where each likelihood reads as
\begin{equation}
-2\log\mathcal{L}_j \propto \sum_{\ell=\ell_{\mathrm{max},j-1}}^{\ell_{\mathrm{max},j}} \sum_{m} \left[\log\det \mathcal{C}^{(j)}_\ell + \left(\bm{a}^{(j)}_{\ell m}\right)^\dagger \left(\mathcal{C}^{(j)}_\ell\right)^{-1} \bm{a}^{(j)}_{\ell m}\right],
\label{eq:gaussian_likelihood_permultipolerange}
\end{equation}
and $\ell_{\mathrm{max},0} \equiv \ell_\mathrm{min} = 2$. The dimension of the data vector and of the covariance matrix depends on the range of multipoles considered, in fact
\begin{equation}
\bm{a}^{(j)}_{\ell m} = 
\left( \begin{matrix}
a^{z_j}_{\ell m} \\
a^{z_{j+1}}_{\ell m} \\
\vdots	\\
a^{z_N}_{\ell m}
\end{matrix} \right),
\qquad \mathcal{C}^{(j)}_\ell = \bm{a}^{(j)}_{\ell m} \left(\bm{a}^{(j)}_{\ell m}\right)^\dagger.
\label{eq:ellmaxvarying_datavector_covariance}
\end{equation}
In equation~\eqref{eq:gaussian_likelihood_permultipolerange} we omit to report explicitly the dependence on cosmological parameters since it is the same of equation~\eqref{eq:gaussian_likelihood_Deltalm}. This decomposition into independent multipole ranges is possible precisely because on linear scales (and for full sky) there is no coupling between different multipoles.

By applying the definition of Fisher matrix (\textit{cf.} equation~\eqref{eq:fisher_matrix_definition}) to the total likelihood we find that the elements of total Fisher matrix are
\begin{equation}
(F_\mathrm{tot})_{\alpha\beta} = F_{\alpha\beta,1} + F_{\alpha\beta,2} + \cdots + F_{\alpha\beta,N},
\label{eq:total_fisher_matrix}
\end{equation}
where $F_{\alpha\beta,j} = \left\langle - \frac{\partial^2\log \mathcal{L}_j}{\partial \theta_{\alpha}\partial \theta_\beta}\right\rangle$. The reasoning presented so far can be easily extended to the case of tracers which are not necessarily galaxies, to the case of multiple tracers and to the case of a Gaussian likelihood for the angular power spectra, reaching identical conclusions.

Finally, in this work we define the maximum multipole in each redshift bin as $\ell_{\mathrm{max},j} \simeq k_\mathrm{max}(z_j) r(z_j)$, where~$k_\mathrm{max}(z_j)$ is the scale where non-linearities become important and $r(z_j)$ is the comoving distance of the redshift bin. We compute~$k_\mathrm{max}(z_j)$ following the typical approach used to introduce corrections to the power spectrum due to non-linearities. However, we stress that a well established procedure does not exist and numerical simulations should be used to assess a more robust~$k_\mathrm{max}(z_j)$. In particular, we want to estimate the scale where the variance of the smoothed linear matter field becomes large enough so that the field cannot be considered linear any more. Such variance is defined as 
\begin{equation}
\sigma^2(R, z) = \int \frac{d^3k}{(2\pi)^3} W^2_R(k, R) P_\mathrm{lin}(k, z), 
\end{equation}
where~$W_R$ is a filter function of characteristic radius~$R$ and~$P_\mathrm{lin}$ is the matter linear power spectrum.

This kind of criterion involves at least two degrees of freedom. The first one is represented by the choice of window function, typically either a top-hat in real space, a Gaussian or a top-hat in Fourier space. The second  is related to the choice of a threshold value~$\sigma_\mathrm{th}$ for the variance: we define the scale where non-linearities become important as $R_\mathrm{max} = k^{-1}_\mathrm{max}$,\footnote{In the literature this quantity is also indicated by $R_\sigma = k^{-1}_\sigma$ or $R_\mathrm{NL} = k^{-1}_\mathrm{NL}$.} where $R_\mathrm{max}$ is the largest smoothing scale such that $\sigma(R_\mathrm{max}, z) = \sigma_\mathrm{th}$. Widely used criteria are $\sigma_{\rm th} = 1$, as in \texttt{Halofit}~\cite{smith:halofit, bird:halofit, takahashi:halofit}, and $\sigma_{\rm th} = \delta_\mathrm{crit.}$, as in \texttt{HMcode}~\cite{mead:hmcode}. Both criteria are typically implemented using a top-hat window function in real space, which is also the standard choice of window function in \texttt{CLASS}. 

\begin{table}
\centerline{
\begin{tabular}{|c|c|c|c|}
\hline
$z_j$ & $R_\mathrm{max}(z_j)$ & $k_\mathrm{max}(z_j)$ & $\ell_{\mathrm{max},j}$\\
\hline
\hline
$0.3$ & $6.7\ \mathrm{Mpc}$ & $0.15\ \mathrm{Mpc}^{-1}$ & $180$ \\
$0.7$ & $4.7\ \mathrm{Mpc}$ & $0.21\ \mathrm{Mpc}^{-1}$ & $550$ \\
$1.1$ & $3.3\ \mathrm{Mpc}$ & $0.31\ \mathrm{Mpc}^{-1}$ & $1100$ \\
$1.5$ & $2.3\ \mathrm{Mpc}$ & $0.43\ \mathrm{Mpc}^{-1}$ & $1900$ \\
$1.9$ & $1.7\ \mathrm{Mpc}$ & $0.59\ \mathrm{Mpc}^{-1}$ & $3000$ \\
\hline
\end{tabular}}
\caption{Non-linear scales and corresponding maximum multipoles for the set of reference mean redshift $z_j = 0.3, 0.7, 1.1, 1.5, 1.9$.}
\label{tab:nonlinear_scale}
\end{table}

Here we adopt $\sigma_\mathrm{th} = 1$ and top-hat in real space window function. The resulting non-linear scales and maximum multipoles are reported in  table~\ref{tab:nonlinear_scale}. This choice of threshold criterion and window function represents a conservative choice, since for larger~$\sigma_\mathrm{th}$ or different window function we would have found a set of larger~$k_\mathrm{max}(z_j)$, thus a set of larger~$\ell_{\mathrm{max},j}$.

More aggressive strategies can be implemented relying on numerical simulations and including a theoretical error to account for uncertainties related to non-linearities, however this analysis goes beyond the scope of this work. Moreover, notice that for a fixed value of the threshold, different choices of window functions provide different values of the field variance, hence different non-linearity scales. This fact does not represent an issue in real numerical implementations, as in \texttt{Halofit} and/or \texttt{HMcode}, since there are other numerical coefficients that are adjusted to match numerical simulation and the model predictions. Therefore $\sigma_\mathrm{th}$ should never be interpreted as an absolute number.

%%%%%%%%%%%%%%%%%%%%%%%%%%%%%%%%%%%%%%%%%%%%%%%%%%%%%%%%%%%%%%%%%%%

\section{Tools and cosmological model specifics}
\label{sec:tools_cosmomodel}

In this section we complete the description of the framework we need to assess whether approximations bias the error estimates. First we present the diagnostic tools used to quantify the effects of the approximations in \S~\ref{subsec:diagnostic_tools}, then in \S~\ref{subsec:setup} we describe the adopted fiducial cosmological model and straw-man survey set-up.

%%%%%%%%%%%%%%%%%%%%%%%%%%%%%%%%%%%%%%%%%%%%%%%%%%%%%%%%%%%%%%%%%%%

\subsection{Diagnostic tools}
\label{subsec:diagnostic_tools}

Establishing how different assumptions or approximations affect the final error estimates requires suitable diagnostic tools. Suppose to have two different Fisher matrices obtained with different assumptions, where we adopt the convention that $F^\mathrm{C}$ (and other quantities with index $^\mathrm{C}$) is obtained with the most correct assumptions, thus $F^\mathrm{I}$ is the incorrect one.

We introduce a matrix the elements of which are given by the ratio of the corresponding elements of $F^\mathrm{C}$ and $F^\mathrm{I}$:
\begin{equation}
\mathcal{R}^\mathrm{Fisher}_{\alpha\beta} = F^\mathrm{C}_{\alpha\beta}/F^\mathrm{I}_{\alpha\beta}.
\label{eq:ratio_fisher}
\end{equation}
The value $\mathcal{R}^\mathrm{Fisher}_{\alpha\beta} \simeq 1$ indicates that the approximations adopted are very good, while deviations from unity flag failures of the adopted approximations. Off-diagonal elements of the $\mathcal{R}^\mathrm{Fisher}_{\alpha\beta}$ matrix can be negative if the signs of $F^\mathrm{C}$ and $F^\mathrm{I}$ are discordant. 

In practical applications, equation~\eqref{eq:ratio_fisher} should not be applied blindly. There might be cases in which, because of the approximation adopted, some elements of $F^\mathrm{I}$ are zero when the corresponding elements of $F^\mathrm{C}$ are not. This can happen, for instance, when the model used to compute $F^\mathrm{C}$ depends on a parameter but the model used for $F^\mathrm{I}$ does not because of the approximation adopted. In these cases it is not appropriate to compare the two full Fisher matrices, but it  is still possible to compare the parts of the Fisher matrices common to both cases.

Equation~\eqref{eq:ratio_fisher} represents a first sanity check. However this is not sufficient to completely assess the impact of the approximation on the inferred parameters. A fair assessment involves the comparison of the two parameters covariance matrices $\Sigma^\mathrm{C}=(F^\mathrm{C})^{-1}$ and $\Sigma^\mathrm{I}=(F^\mathrm{I})^{-1}$. Therefore we introduce the ratio of the two covariance matrices~$\mathcal{R}^{\rm Covar.}$, the elements of which are given by
\begin{equation}
\mathcal{R}^\mathrm{Covar.}_{\alpha\beta} = \Sigma^\mathrm{C}_{\alpha\beta} / \Sigma^\mathrm{I}_{\alpha\beta}.
\label{eq:ratio_covariance}
\end{equation}
In fact, even if individual elements of the two Fisher matrices are similar, specific elements of their inverses might not, due to the matrix inversion operation. The effect of the approximation(s) on the parameter marginalised errors, $\sigma_{\theta_\alpha}$,  can be evaluated by considering the diagonal elements
\begin{equation}
\sqrt{\mathcal{R}^\mathrm{Covar.}_{\alpha\alpha}} = \sigma^\mathrm{C}_{\theta_\alpha} / \sigma^\mathrm{I}_{\theta_\alpha},
\end{equation}
where $\sigma^\mathrm{C,I}_{\theta_\alpha} = \sqrt{\Sigma^\mathrm{C,I}_{\alpha\alpha}}$. As above, values of this ratio close to unity correspond to small differences in the modelling (i.e., good approximations). Notice that both statistics, equations~\eqref{eq:ratio_fisher} and~\eqref{eq:ratio_covariance}, are independent of the~$f_\mathrm{sky}$ parameter, provided that it is the same for~$F^\mathrm{C}$ and~$F^\mathrm{I}$.

Degeneracies between parameters are typically visualised using confidence ellipses. The confidence ellipses are drawn in parameter space starting from the parameter covariance matrix $\Sigma_{\alpha\beta}$ and are given by
\begin{equation}
\left(\frac{\Delta\theta_\alpha}{\sigma_{\theta_\alpha}}\right)^2 - 2\rho_{\alpha\beta}\frac{\Delta\theta_\alpha}{\sigma_{\theta_\alpha}}\frac{\Delta\theta_\beta}{\sigma_{\theta_\beta}} +\left(\frac{\Delta\theta_\alpha}{\sigma_{\theta_\beta}}\right)^2 = \left(1-\rho^2_{\alpha\beta}\right)\Delta\chi^2,
\label{eq:ellipses}
\end{equation}
where~$\Delta\theta_\alpha=\theta_\alpha-\theta_\alpha^\mathrm{fid}$ is the distance of the parameter~$\theta_\alpha$ from its fiducial value~$\theta_\alpha^\mathrm{fid}$, $\rho_{\alpha\beta}~=~\Sigma_{\alpha\beta}/\sqrt{\Sigma_{\alpha\alpha}\Sigma_{\beta\beta}}$ is the correlation coefficient between the $\alpha$-th and $\beta$-th parameters, and assuming a bivariate Gaussian distribution for couples of parameters we can compute $\Delta\chi^2 = -2\log\left(1-\mathrm{CL}\right)$, where $\mathrm{CL}$ is the desired confidence level. Confidence levels of $\mathrm{CL} = (0.683, 0.954, 0.997)$ correspond to $\Delta\chi^2 = (2.30, 6.16, 11.62)$. In this case smaller values of the sky coverage $f_\mathrm{sky}$ correspond to a widening of the ellipses.

Figure-of-Merit (FoM) are useful to compare different experiments. The inverse of the area of the ellipse associated to the $95\%\ \mathrm{CL}$ for given couples of parameters $(\theta_\alpha,\theta_\beta)$, with $\alpha\neq\beta$ is  given by
\begin{equation}
\mathrm{FoM}_{\alpha\beta} = \frac{\pi}{A^\mathrm{ellipse}_{\alpha\beta}} = \frac{1}{\Delta\chi^2\sigma_{\theta_\alpha}\sigma_{\theta_\beta}\sqrt{1-\rho^2_{\alpha\beta}}}
\end{equation}
and in principle it depends on the coverage of the sky, however the ratio of FoM computed with different covariance matrices 
\begin{equation}
\mathcal{R}^\mathrm{FoM}_{\alpha\beta} = \mathrm{FoM}^\mathrm{C}_{\alpha\beta} / \mathrm{FoM}^\mathrm{I}_{\alpha\beta}
\end{equation}
is independent of $f_\mathrm{sky}$ and also of the chosen confidence level, $\mathrm{CL}$.

Most of our diagnostic tools are represented by ratios~$\mathcal{R}_{\alpha\beta}$ of the same quantity computed under different assumptions. For the sake of clarity, the figures of the next sections will always use the same color bar: the colors blue, white and red indicate, respectively,  ratios larger than unity ($\mathcal{R}_{\alpha\beta}>1$), unity ($\mathcal{R}_{\alpha\beta}=1$) and smaller than unity ($\mathcal{R}_{\alpha\beta}<1$). Therefore, the color blue indicates that entries obtained with the approximation have been underestimated, the color red indicates they have been overestimated.

%%%%%%%%%%%%%%%%%%%%%%%%%%%%%%%%%%%%%%%%%%%%%%%%%%%%%%%%%%%%%%%%%%%

\subsection{Cosmological model and straw-man galaxy survey set-up}
\label{subsec:setup}

We consider a $\Lambda\mathrm{CDM}+f_\mathrm{NL}$ model, where the standard $\Lambda\mathrm{CDM}$ cosmological model is extended to include the contribution of primordial non-Gaussinities of the local type to the total galaxy bias. The set of cosmological parameters $\left\lbrace\theta_\alpha\right\rbrace$  is
\begin{equation}
\left\lbrace\theta_\alpha\right\rbrace = \left\lbrace h, \omega_\mathrm{b}, \omega_\mathrm{cdm}, n_\mathrm{s}, \{b_\mathrm{g}\}, f_\mathrm{NL} \right\rbrace,
\label{eq:fisher_parameters} 
\end{equation}
where $h$ is the present-day reduced Hubble expansion rate, $\omega_\mathrm{b}$ is the present-day physical baryon density, $\omega_\mathrm{cdm}$~is the present-day physical cold dark matter density, $n_\mathrm{s}$ is the scalar spectral index, and $\{b_\mathrm{g}\}$ and $f_\mathrm{NL}$ are the parameters that enter in our definition of the total bias in equation~\eqref{eq:total_galaxy_bias}. Note that we have included a set of scale- and redshift-independent Gaussian galaxy biases $\{b_\mathrm{g}\}$ to account also for the multiple tracers case. \footnote{In the single tracer case, the fiducial value of cosmological parameters of interests reads as
\begin{equation*}
\left\lbrace h, \omega_\mathrm{b}, \omega_\mathrm{cdm}, n_\mathrm{s} \right\rbrace = \left\lbrace 0.6727, 0.02225, 0.1198, 0.9645 \right\rbrace .
\end{equation*}
Furthermore, we use $\log 10^{10}A_\mathrm{s}=3.0940$, three massive neutrinos with $m_\nu=0.02\ \mathrm{eV}$. For the single tracer case we choose $\left\lbrace b_X, f_\mathrm{NL} \right\rbrace = \left\lbrace 2.0, 0.0 \right\rbrace$, while for the multiple tracers case we used $\left\lbrace b^\mathrm{Eu-l}_\mathrm{g}, b^\mathrm{Sp-l}_\mathrm{g}, f_\mathrm{NL} \right\rbrace = \left\lbrace 2.0, 1.4, 0.0 \right\rbrace$.} We do not include the amplitude of the primordial scalar perturbations~$A_\mathrm{s}$ in the Fisher analysis since we are presenting a proof of principle example. We are aware of the importance of including this parameter in a real analysis, especially because it is partially degenerate with the amplitude of the total galaxy bias. We consider five scale- and redshift-independent values of the galaxy magnification bias parameter, given by $s_\mathrm{g}=\{0.0,0.2,0.4,0.6,0.8\}$ and chosen symmetrically around the value $s_\mathrm{g}=0.4$, which corresponds to a vanishing lensing contribution (see also appendix~\ref{app:relativistic_number_counts}).\footnote{The range of magnification bias parameter values is representative of plausible scenarios for future large-scale structure surveys, see e.g., appendix A of ref.~\cite{jeliccizmek:magbiassurveys}. We are aware of the importance of including also the magnification bias evolution in redshift, however since the purpose of this paper is not making forecasts for specifics surveys, we chose to adopt a simpler prescription. We stress that uncertainties in the modelling of bias, magnification bias and evolution bias parameters are a possible source of error mis-estimation. The effects of approximations in more realistic scenarios can be extrapolated from the results we report in the following sections.}

In order to illustrate the impact of survey-dependent specifications, we consider three straw-man surveys. The first has a uniform galaxy distribution with $d^2N_\mathrm{g}/dzd\Omega = 1070\ \mathrm{gal/deg}^2$, while the other two have a redshift distribution parametrised as 
\begin{equation}
\frac{d^2N_\mathrm{g}}{dzd\Omega} = \mathcal{A} \left(\frac{z}{z_0}\right)^\alpha e^{-\left(z/z_0\right)^{\beta}}.
\label{eq:tracer_distribution}
\end{equation}
The second straw-man survey is inspired by the Euclid galaxy distribution at redshift $z>0.9$ ($\mathcal{A}=2400\ \mathrm{gal/deg^2},\ z_0=0.54,\ \alpha=4,\ \beta=1.5$)\footnote{These values correspond to the case reported in column 4 of table 3 of ref.~\cite{amendola:euclidwhitepaperI} and are compatible with the galaxy number density reported in ref.~\cite{blanchard:euclidforecasts}.}~\cite{laureijs:euclidsciencebook, amendola:euclidwhitepaperI, blanchard:euclidforecasts}, while the third one is inspired by a SPHEREx-like galaxy population ($\mathcal{A}=29300\ \mathrm{gal/deg^2},\ z_0=0.53,\ \alpha=1.1,\ \beta=1.5$)\footnote{These values refer to the SPHEREx's $\sigma(z)/(1+z)<0.1$ sample.}~\cite{dore:spherexwhitepaperI, dore:spherexwhitepaperII}. We denote the last two straw-man surveys as ``Euclid-like'' and ``SPHEREx-like'', with corresponding superscripts ``Eu-l'' and ``Sp-l''. The uniform and Euclid-like populations are used for the single tracer cases treated in sections~\ref{sec:likelihood_modeling} and~\ref{sec:signal_modeling}, and they share the total number of observed galaxies in the redshift range $z\in [0.1,2.1]$. The SPHEREx-like population is used along with the Euclid-like one for the multiple tracer case.

We choose top-hat window functions $W(z,z_i,\Delta z_i)$ in all the cases, as in spectroscopic galaxy surveys. In the following we use the set of mean redshift~$\{z_i\}=\{0.3, 0.7, 1.1, 1.5, 1.9\}$. We check different redshift binning choices: we consider the case with $\Delta z=0.2$ and $\Delta z=0.3$, corresponding to the non-overlapping and overlapping redshift bins case, respectively. In the latter case the galaxy populations are defined in the redshift range $z\in [0.0,2.2]$. We refer the interested reader to ref.~\cite{bailoni:fishermatriximprovement} for a more extensive analysis on the effects of different window functions and redshift binnings. 

Unless otherwise indicated, the maximum multipoles used in each redshift bin are reported in table~\ref{tab:nonlinear_scale}: this choice ensures that only linear scales are included over the entire redshift range considered. In sections~\ref{sec:signal_modeling} and~\ref{sec:multitracing} we present results varying the maximum multipole range following the method discussed in \S~\ref{subsec:sky_maps_analysis}.

%%%%%%%%%%%%%%%%%%%%%%%%%%%%%%%%%%%%%%%%%%%%%%%%%%%%%%%%%%%%%%%%%%%

\section{Effects of approximations in the likelihood: neglecting covariance between redshift bins}
\label{sec:likelihood_modeling}

One approximation which is often adopted is to consider only the diagonal part of the~$\mathcal{M}_\ell$ covariance matrix defined in equation~\eqref{eq:Mell_covariance_matrix}, i.e., to neglect partially the covariance between different redshift bins. This approximation changes the shape of the likelihood: both Fisher matrices in equations~\eqref{eq:fisher_likelihood_delta} and~\eqref{eq:fisher_likelihood_cl} describe accurately the curvature around the likelihood maximum if and only if all elements of the respective covariance matrices are included. We expect this effect to be more pronounced when the correlation between redshift bins is higher, for instance when redshift bins overlap. 

\begin{figure}
\centerline{
\includegraphics[width=1.0\columnwidth]{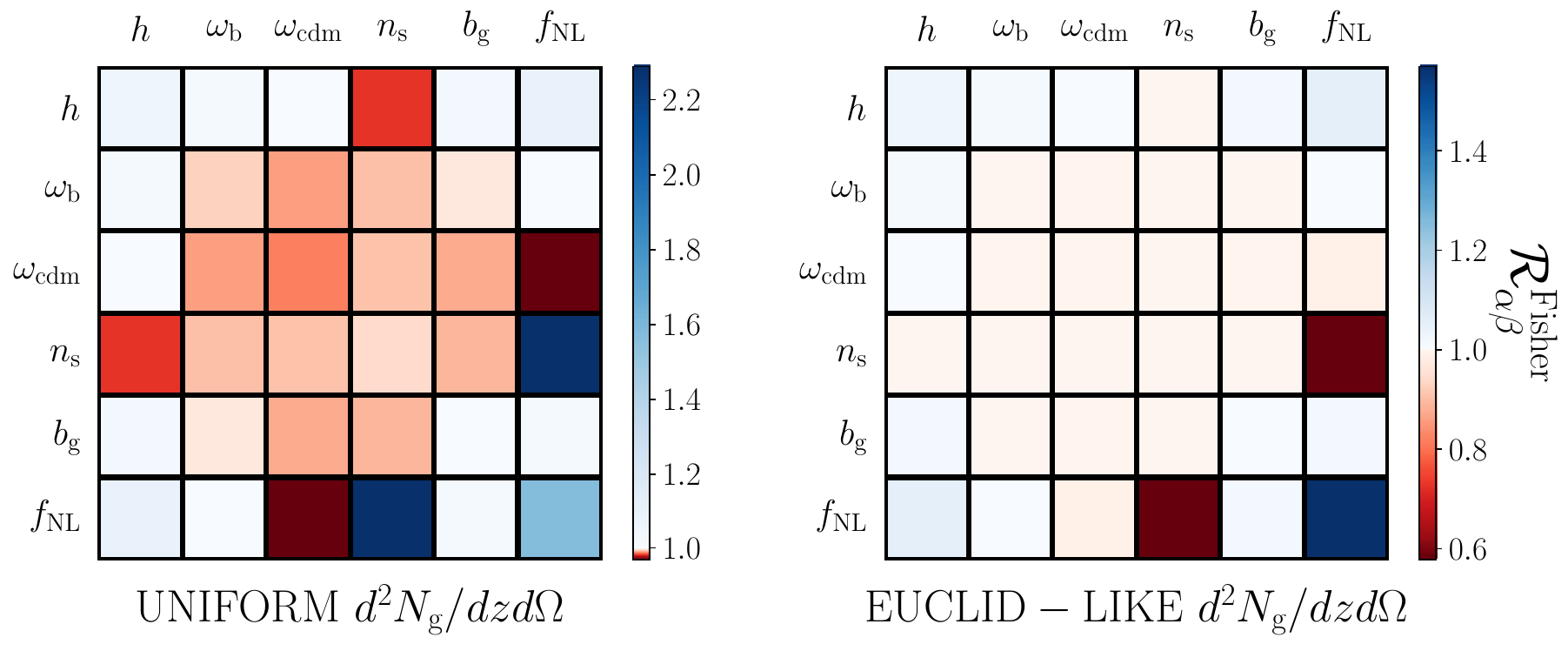}}
\caption{Ratio of the Fisher matrix elements obtained including and neglecting the off-diagonal terms of the covariance matrix (i.e., neglecting correlations between redshift bins), assuming $s_\mathrm{g}=0.0$ and non-overlapping redshift bins. We show results for a uniform (\textit{left panel}) and an Euclid-like (\textit{right panel}) galaxy population. Note the different ranges between the two panels.} 
\label{fig:ratio_good_vs_bad_covariance_matrix}
\end{figure}

We show 	the ratio between the Fisher matrix elements $\mathcal{R}^\mathrm{Fisher}_{\alpha\beta}$ computed with the complete data covariance matrix $\mathcal{M}_\ell$ and computed only with its diagonal part in figure~\ref{fig:ratio_good_vs_bad_covariance_matrix}.  Results as shown for the uniform and Euclid-like galaxy populations, assuming~$s_\mathrm{g}=0.0$ as galaxy magnification bias parameter and non-overlapping redshift bins. The Fisher matrix elements relative to the standard cosmological parameters are affected by the approximation only at the few percent level. However, the ratio of some of the elements related to the primordial non-Gaussianity parameter is significantly different from unity, especially elements involving $n_{\rm s}$ and $f_{\rm NL}$. The ratio of marginalised errors $\sqrt{\mathcal{R}^\mathrm{Covar.}_{\alpha\alpha}}$ indicates that this approximation overestimates real errors on $f_\mathrm{NL}$ by $20-30\%$, whereas the errors on the other parameters are almost unchanged. Although we show only the $s_\mathrm{g}=0.0$ case,  results  for other values of the magnification bias parameter are very similar.

\begin{figure}
\centerline{
\includegraphics[width=1.0\columnwidth]{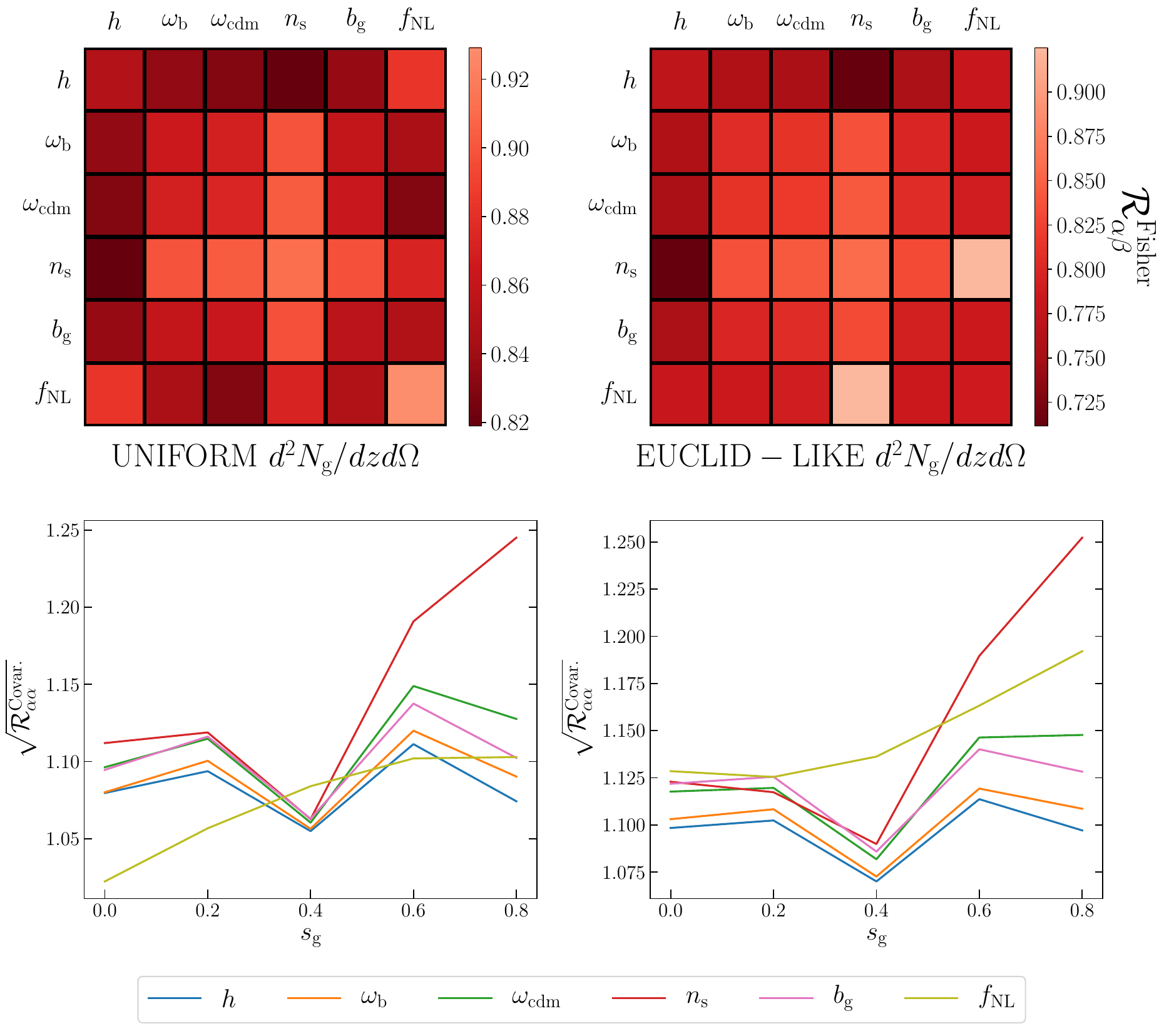}}
\caption{\textit{Top panels:} Ratio of the Fisher matrix elements obtained including and neglecting the off-diagonal terms of the covariance matrix, assuming $s_\mathrm{g}=0.0$ and overlapping redshift bins. \textit{Bottom panels}: marginalised errors ratio $\sqrt{\mathcal{R}^\mathrm{Covar.}_{\alpha\alpha}}$ for different values of the magnification bias parameter $s_\mathrm{g}$. We present the results obtained for a uniform and an Euclid-like galaxy population in the \textit{left} and \textit{right panels}, respectively.} 
\label{fig:ratio_good_vs_bad_covariance_matrix_overlapping_bins}
\end{figure}

In the case of overlapping redshift bins, the effects are, unsurprisingly, more significant, as shown in figure~\ref{fig:ratio_good_vs_bad_covariance_matrix_overlapping_bins}. Fisher matrix elements relative to all parameters are significantly affected at the $10\%$ to $30\%$ level. These changes affect the marginalised parameters errors in ways hard to determine a priori as illustrated in the lower panels of the same figure. However, in this specific set up, errors obtained when using the approximation are underestimated and this conclusion holds both for the uniform and Euclid-like galaxy populations, but the magnitude of the effect depend somewhat on the adopted fiducial value of the magnification bias parameter $s_\mathrm{g}$, as it can be appreciated from the bottom panels of figure~\ref{fig:ratio_good_vs_bad_covariance_matrix_overlapping_bins}. In the $s_\mathrm{g}=0.4$ case the effect of this approximation is less significant because there are no lensing effects that correlate tracers in different redshift bins except for local effects due to overlapping volumes between redshift bins. We refer the reader to \S~\ref{subsec:cosmic_magnification} for a more in depth discussion on cosmic magnification. Although in this specific example degeneracies between parameters are not affected much by the approximation adopted, we caution the reader that they might be affected in other set-ups, since this effect is strongly case-dependent.

Although we have presented a specific example with specific galaxy redshift surveys and using their angular power spectrum as the summary statistics, qualitatively the results will hold independently of the specific observable or data set used, e.g., secondary effects on the cosmic microwave background, cosmic shear, and so on. The effect arises because the approximation of neglecting correlations among different redshift bins (chief among them magnification) induces an incorrect shape of the likelihood. Similarly, using a Gaussian likelihood approximation where the likelihood is in fact non-Gaussian will also induce mis-estimation of the errors, see e.g., ref.~\cite{repp:gaussianassumption}.

%%%%%%%%%%%%%%%%%%%%%%%%%%%%%%%%%%%%%%%%%%%%%%%%%%%%%%%%%%%%%%%%%%%

\section{Effects of approximations in the modelling of the observable}
\label{sec:signal_modeling}

Approximations in the computation and modelling of the target observable (here the angular power spectrum) can have subtle effects on the error estimate. For example,  at large scales (low multipoles) and in absence of systematics, the error on the signal is dominated by cosmic variance and its magnitude depends on the signal itself, i.e., $\sigma_{C_\ell}=\sqrt{2/(2\ell+1)}C_\ell$. Therefore, not only the physical signal, but also its covariance is affected by approximations.

As can be seen in equation~\eqref{eq:numbercount_fluctuation} and in appendix~\ref{app:relativistic_number_counts}, a large number of different physical effects contribute to the total signal. Not all of them depend on the same parameters in the same way; for instance, the total bias~$b_{X,\mathrm{tot}}$ enters only in the density contribution while the magnification bias parameter affects primarily to the lensing contribution and does not affect the density term. For this reason, it could be tempting (and it is often done) not to include contributions that do not depend on the parameters of interest. One example is neglecting the lensing contribution in equation~\eqref{eq:numbercount_fluctuation} when studying primordial non-Gaussianities, given that non-Gaussianities affect only the galaxy bias, hence only the density contribution. We illustrate a specific example in subsection~\ref{subsec:cosmic_magnification} where we show the effect of neglecting cosmic magnification. Cosmic magnification does not depend on (i.e., has zero derivative with respect to) the parameters entering in the total bias definition ($b_X$ and~$f_\mathrm{NL}$). This condition is necessary, but it is not sufficient, to ignore the contribution completely in the error forecast: cosmic magnification changes the signal, thus the covariance matrices that enter in equations~\eqref{eq:fisher_likelihood_delta} and~\eqref{eq:fisher_likelihood_cl}.

Even when all the contributions are included in the theoretical modeling of the observable, commonly used numerical approximations might still be insufficient. As an example, we show how the widely used Limber approximation~\cite{limber:approximation} effectively changes the shape of the signal in subsection~\ref{subsec:limber_approximation}. In turn, this change affects the likelihood and therefore marginal errors and correlations among parameters. While this type of approximations may be time-saving, they should not be used without first assessing  very carefully their impact on the analysis. Notice that Limber approximation is just one of the numerous approximations that are usually taken both at large and small scales. For example, we refer the interested reader to ref.~\cite{reimberg:halofit}, where the authors show that similar effects are present at small, non-linear scales when using the analytical fit to the power spectrum provided by \texttt{Halofit}.

%%%%%%%%%%%%%%%%%%%%%%%%%%%%%%%%%%%%%%%%%%%%%%%%%%%%%%%%%%%%%%%%%%%

\subsection{Effects of cosmic magnification}
\label{subsec:cosmic_magnification}
Magnification lensing changes the sources surface density on the sky (see section~\ref{sec:observable_modeling_fisher_matrix}) and it is sensitive to the full matter distribution~\cite{turner:magnificationbias, kaiser:limberapproximation, matsubara:lensing, hui:magnificationbias, loverde:cosmicmagnification}. Since it does not depend on the tracers' bias, the lensing contribution is often neglected in the study and forecasts of primordial non-Gaussianity, which signal appears in the total galaxy bias at large scales. 
 
\begin{figure}
\centerline{
\includegraphics[width=1.0\columnwidth]{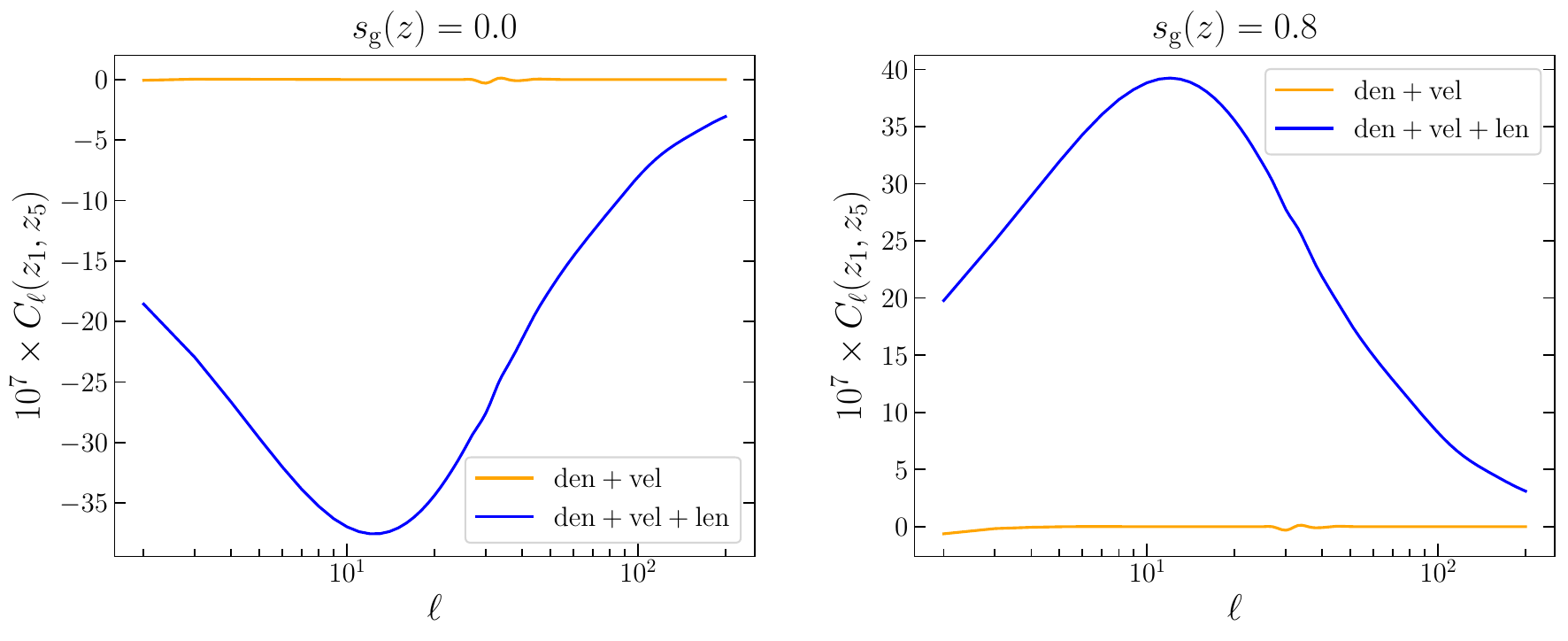}}
\caption{Cross-bin ($0.1<z_1<0.5$ and $1.7<z_5<2.1$) angular power spectra with (orange curve) and without (blue curve) lensing contribution, for $s_\mathrm{g}=0.0$ (\textit{left panel}), and $s_\mathrm{g}=0.8$ (\textit{right panel}).}
\label{fig:cl_cross}
\end{figure} 
 
The parameters $b_\mathrm{g}$ and $f_\mathrm{NL}$ enter only in the intrinsic clustering term~$\Delta^\mathrm{den}_\ell$ (see appendix~\ref{app:relativistic_number_counts}), therefore the naive expectation is that excluding the velocity~$\Delta^\mathrm{vel}_\ell$ or lensing~$\Delta^\mathrm{len}_\ell$ contributions does not affect the final error estimate. For instance, some velocity terms, at linear level, have the same $k^{-2}$ scale dependence as the non-Gaussian halo bias, hence they act as an effective $f_\mathrm{NL}$~\cite{raccanelli:dopplerterms}. Failing to include them in the theoretical modelling  will bias the estimate of primordial non-Gaussianity, $f_{\rm NL}$, and  lead to an incorrect estimate of the parameter errors. A correct implementation of the velocity terms has been shown to be crucial not to bias parameter estimation~\cite{camera:probingpng, tanidis:includingvelocityterms, tanidis:includinglensing}.

\begin{figure}
\centerline{
\includegraphics[width=1.0\columnwidth]{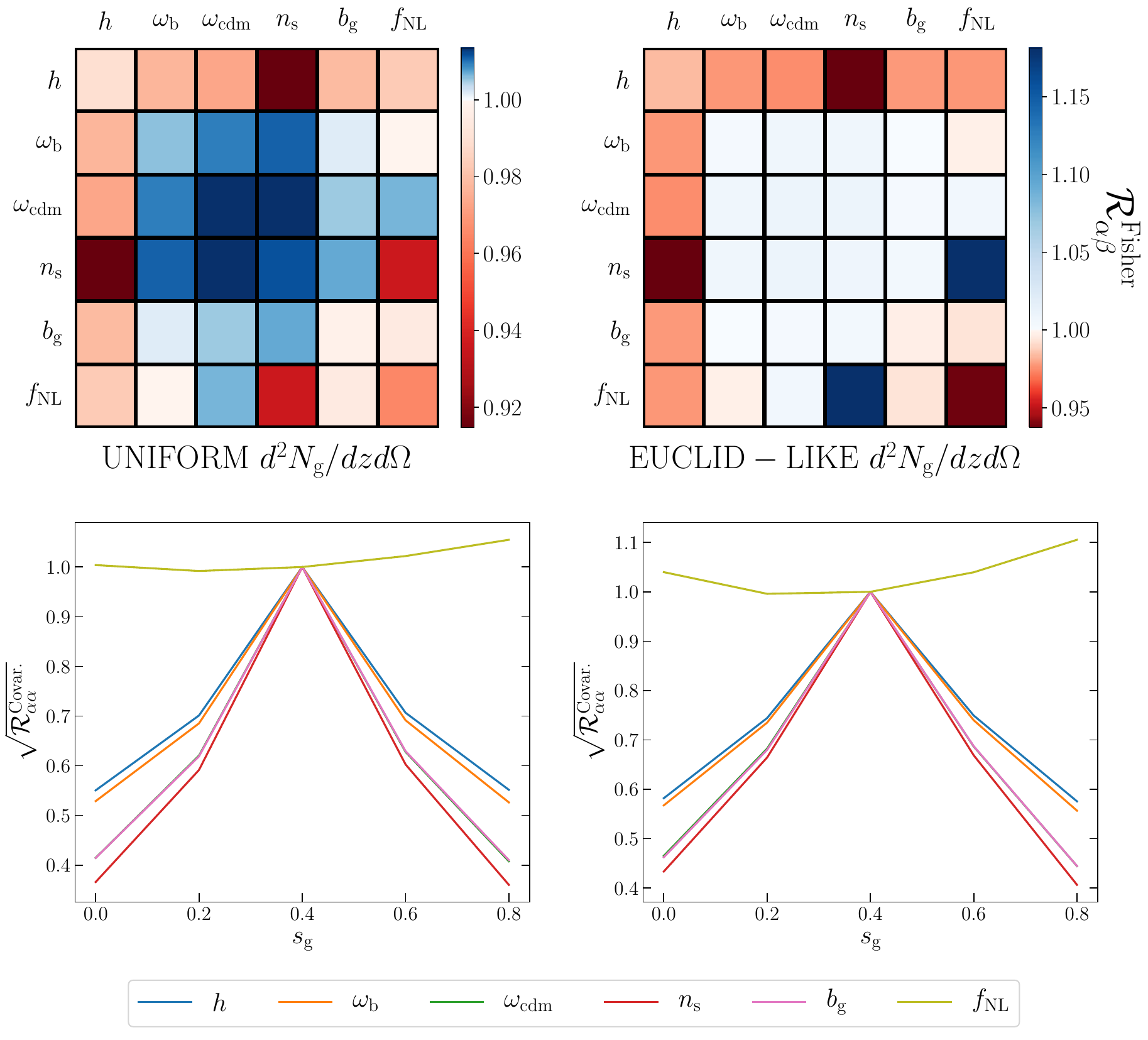}}
\caption{\textit{Top panels:} Ratio of the Fisher matrix elements obtained including and neglecting the lensing contribution in the galaxy angular power spectrum, for $s_\mathrm{g}=0.6$. \textit{Bottom panels}: marginalised errors ratio $\sqrt{\mathcal{R}^\mathrm{Covar.}_{\alpha\alpha}}$ for different values of the magnification bias parameter. We present the results obtained for a uniform and an Euclid-like galaxy population with non-overlapping redshift bins in the \textit{left} and \textit{right panels}, respectively.} 
\label{fig:ratio_lensing_vs_nolensing_covariance_matrix}
\end{figure}

The lensing contribution is typically subdominant in the angular power spectrum of sources in the same redshift bin. However, it dominates the signal in the cross-bin correlation. In particular, the dominant term is the density-lensing contribution $\Delta^\mathrm{den}_\ell\Delta^\mathrm{len}_\ell \propto b_{X,\mathrm{tot}}(2-5s_X)$, which represents the observed correlation between foreground and background galaxies due to gravitational lensing and  depends on the clustering properties of the tracers; hence, it is affected by primordial non-Gaussianity of the local type. As for all contributions that include~$\Delta^\mathrm{len}_\ell$, this term vanishes for $s_X=0.4$.

Figure~\ref{fig:cl_cross} illustrates the  magnitude of the lensing effects in the galaxy-galaxy cross-z-bin angular power spectrum, for significantly separated redshift bins ($0.1<z_1<0.5$ and $1.7<z_5<2.1$). We show the angular power spectrum for two different values of the magnification bias, $s_\mathrm{g}=\{0.0,0.8\}$, for which cosmic magnification increases and decreases the number of observed objects behind the lens, respectively. As can be seen from the figure, the angular power spectra between bins with large radial separation change by orders of magnitude when lensing effects are included. Therefore, neglecting cosmic magnification heavily impacts the magnitude of covariance matrix elements used in the Fisher analysis, so that all cosmological parameters are affected. Moreover, lensing effects help to break degeneracies between parameters, for instance between the amplitude of scalar perturbations $A_\mathrm{s}$ and the galaxy bias $b_\mathrm{g}$.   

Figure  \ref{fig:ratio_lensing_vs_nolensing_covariance_matrix} illustrates the effects of neglecting the lensing effects  on the Fisher matrix elements and on the marginalised parameters errors. While this approximation affects the Fisher matrix elements only at the $10\%$ level, it causes an overestimate of all the marginalised parameters errors (except for $f_{\rm NL}$) by up to~$40-60\%$ for~$s_\mathrm{g}=0.0$ and~$s_\mathrm{g}=0.8$. The overestimate is of order~$30\%$ also for the less ``extreme'' values of~$s_\mathrm{g}=0.2$ and~$0.6$, highlighting the fact that the estimate of the marginalised error is very sensitive to lensing effects. Although neglecting lensing magnification does not significantly impact the error on $f_{\rm NL}$, mis-estimating the errors for the rest of the parameters affects parameter estimation when different experiments with different parameter degeneracies are combined. This example illustrates that the marginalised error estimate is really sensitive to the values of \textit{all} Fisher matrix elements, hence comparison between different approximations must be performed on the full Fisher matrix not on selected elements (or alternatively on the parameters covariance matrix). This result is valid for both uniform and Euclid-like galaxy populations and for different values of the magnification bias.

Only in the case where the lensing contribution is negligible, i.e., for~$s_\mathrm{g} \simeq 0.4$, the approximation holds; however $s_\mathrm{g}$ is usually a poorly known quantity and therefore assuming~$s_\mathrm{g}\equiv 0.4$ may not be justified. In practice, the overall importance of lensing effects also depends on the redshift-dependence of the magnification bias parameter for the selected galaxy population, on the galaxy redshift distribution and on the selected redshift binning. A more realistic treatment of these issues can be found, e.g., in refs.~\cite{bernal:emu, bacon:skaredbook}.

\begin{figure}
\centerline{
\includegraphics[width=1.0\columnwidth]{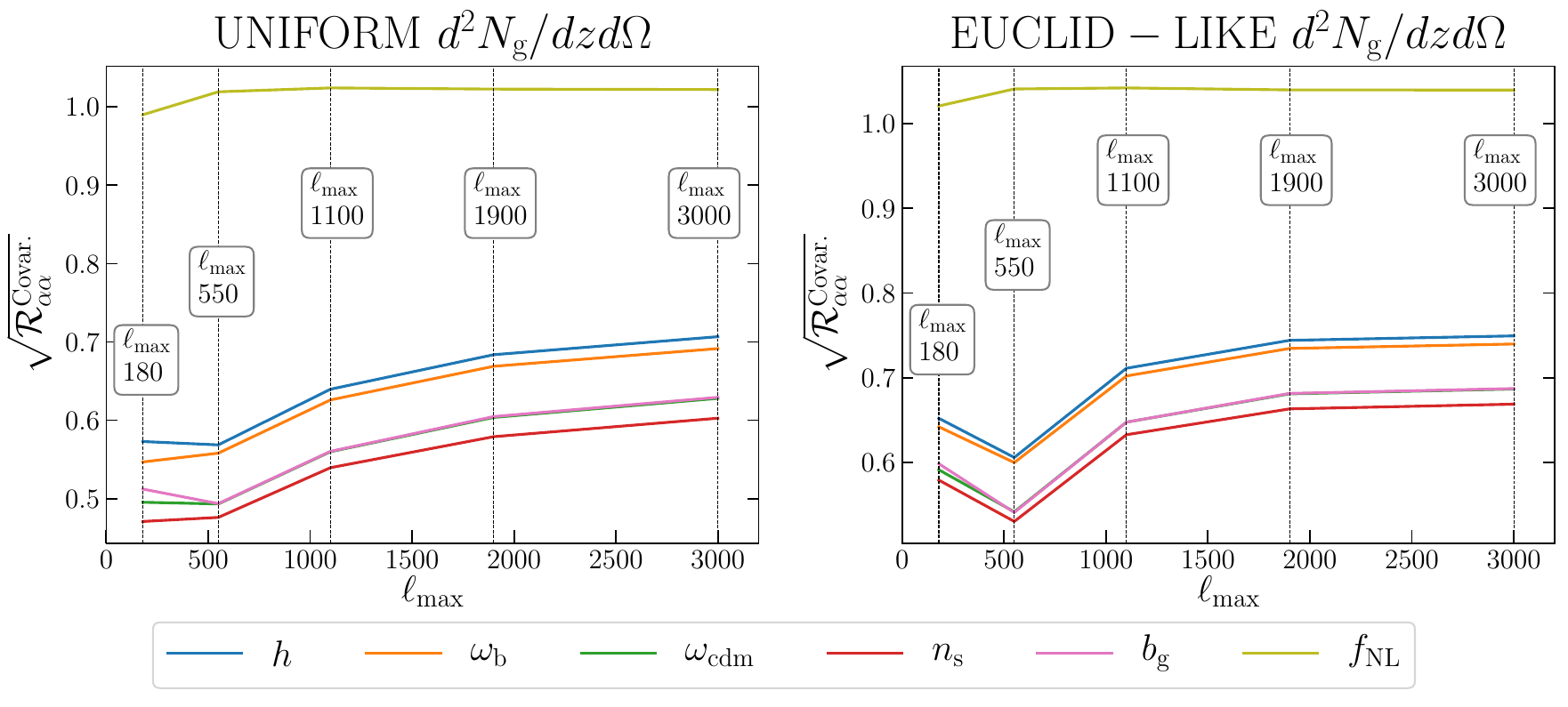}}
\caption{Evolution of the marginalized error ratio $\sqrt{\mathcal{R}^\mathrm{Covar.}_{\alpha\alpha}}$ as a function of the maximum multipole range chosen in the analysis for a uniform and an Euclid-like galaxy population with non-overlapping redshift bins in the \textit{left} and \textit{right panels}, respectively, when $s_\mathrm{g}=0.6$.} 
\label{fig:sigma_ratio_lensing_vs_nolensing}
\end{figure}

We should point out  that the results presented in figure~\ref{fig:ratio_lensing_vs_nolensing_covariance_matrix} are mitigated by our choice of maximum multipole range. We show in figure~\ref{fig:sigma_ratio_lensing_vs_nolensing} how the mismatch between accurate and non-accurate errors depends on the maximum multipole range used in the analysis for the~$s_\mathrm{g}=0.6$ case. This maximum multipole range analysis follows the procedure described in \S~\ref{subsec:sky_maps_analysis}. We notice that for both galaxy populations the marginalized error ratio~$\sqrt{\mathcal{R}^\mathrm{Covar.}_{\alpha\alpha}}$ ``saturates'' when all the multipole ranges are included. However, it never reaches the value~$\sqrt{\mathcal{R}^\mathrm{Covar.}_{\alpha\alpha}}=1$, hence increasing the multipole range never fully  corrects for this systematic effect. Therefore, in this specific case, a conservative analysis is more affected by this approximation than an aggressive one.

This behaviour is expected: increasing the maximum multipole means increasing the number of multipole ranges, hence it corresponds to include more terms in equation~\eqref{eq:total_fisher_matrix}. However the difference between the accurate and non-accurate extra terms are expected to decrease at high $\ell$. Each multipole range that we include involves a smaller number of redshift bins. Therefore, we are cross-correlating redshift bins that are less separated, i.e., that are less affected by lensing effects. Since the cross-bin angular power spectra typically enters in the off-diagonal terms of the covariance matrix, the case where we neglect cosmic magnification presents similarities with the case of section~\ref{sec:likelihood_modeling} where we neglect off-diagonal terms. We checked that also for the cases of section~\ref{sec:likelihood_modeling} we observe a trend similar to that reported in figure~\ref{fig:sigma_ratio_lensing_vs_nolensing}.

Finally, we note that the same analysis should be done also for the gravity contribution of equation~\eqref{eq:numbercount_fluctuation}. These effects are relevant at scales close to the horizon, i.e., at low multipoles, where primordial non-Gaussianities might contribute significantly. We leave this analysis for future work.

%%%%%%%%%%%%%%%%%%%%%%%%%%%%%%%%%%%%%%%%%%%%%%%%%%%%%%%%%%%%%%%%%%%

\subsection{Effects of Limber approximation}
\label{subsec:limber_approximation}

\begin{figure}
\centerline{
\includegraphics[width=1.0\columnwidth]{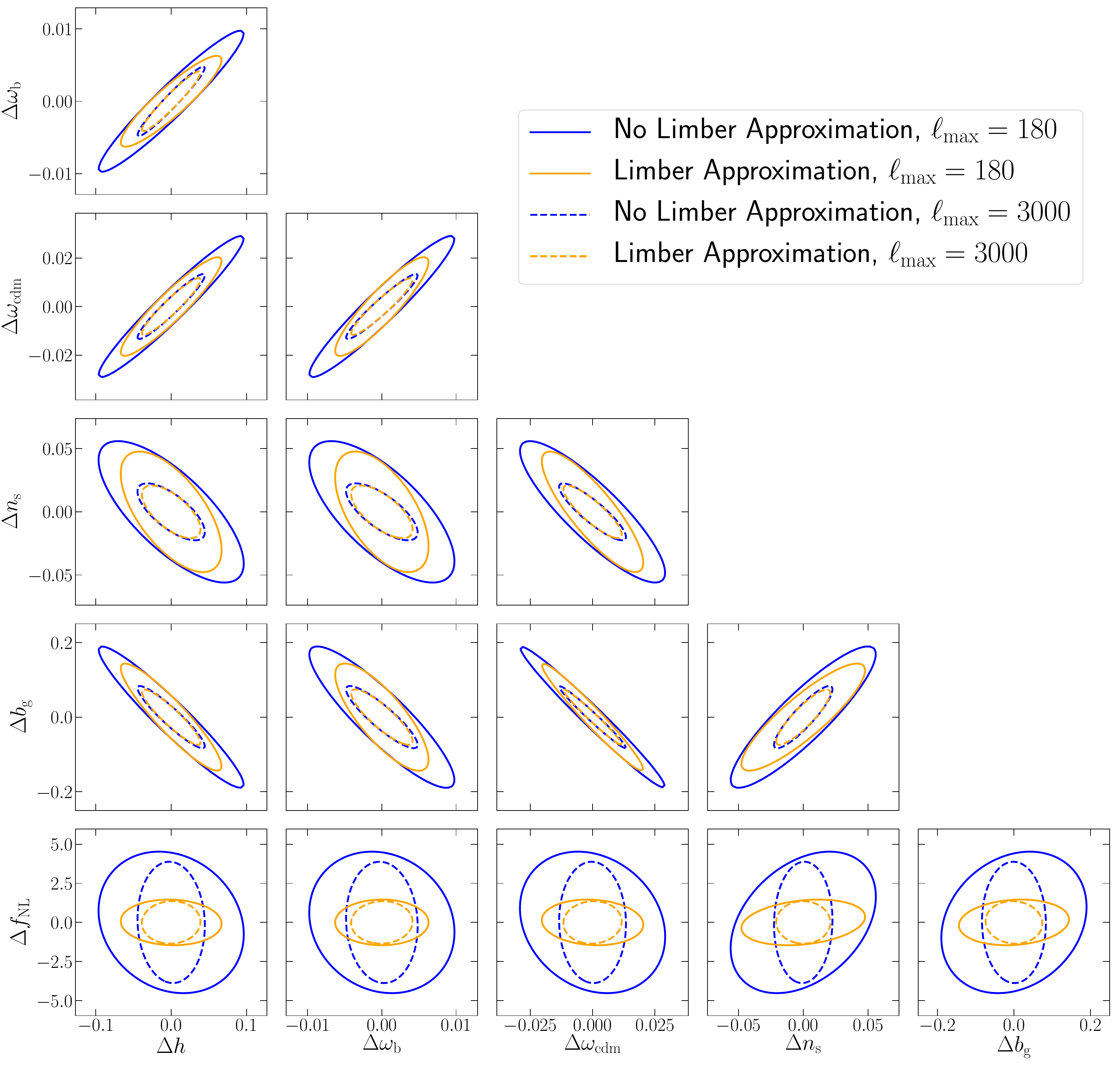}}
\caption{Two-dimensional marginalised $68\%$ CL regions for the Euclid-like galaxy population with non-overlapping redshift bins, using (orange line) or not using (blue line) the Limber approximation. We show constraints for the $s_\mathrm{g}=0.6$ case and two different maximum multipole, $\ell_\mathrm{max}=180$ and~$\ell_\mathrm{max}=3000$. Ellipses, drawn according to equation~\eqref{eq:ellipses}, are centred around the fiducial values of the parameters.} 
\label{fig:confidence_ellipses_limber}
\end{figure}

The Limber approximation (pioneered in cosmology in ref.~\cite{kaiser:limberapproximation}) is used in the limit of small radial and angular separation between galaxies, and it is particularly useful as it simplifies the calculation of angular power spectra in presence of highly oscillatory spherical Bessel functions. Until recently, this limit was almost correct, as most galaxy surveys observed small patches of the sky and they were not very deep. However, forthcoming and future surveys will cover larger patches and go to higher redshift, therefore an accurate ``wide-angle'' treatment of the curvature of the sky and of radial separation must be used to model galaxy clustering.

The Limber approximation substitutes spherical Bessel functions with Dirac delta functions, $j_\ell(kr)\simeq\sqrt{\frac{\pi}{2\ell+1}}\delta^D\left(\ell+\frac{1}{2}-kr\right)$, introducing an error of order $\mathcal{O}(1/\ell)$~\cite{loverde:extendedlimber} which becomes negligible at high multipoles ($\ell\gg 10$). Even though there are methods to accurately compute such oscillatory integrals in a fast and accurate way, see e.g., refs.~\cite{gebhardt:2fast, shoneberg:clcomputation}, the Limber approximation remains a widely used tool, even at large scales (small $\ell$) where its accuracy drops~\cite{simon:limberaccuracy, kitching:limberaccuracy}. Although the impact of this approximation in current data analysis of weak-lensing and cosmic shear has been found to be subdominant~\cite{kilbinger:limberweaklensing, lemos:limberweaklensing}, this will not be the case for future surveys.

\begin{figure}
\centerline{
\includegraphics[width=1.0\columnwidth]{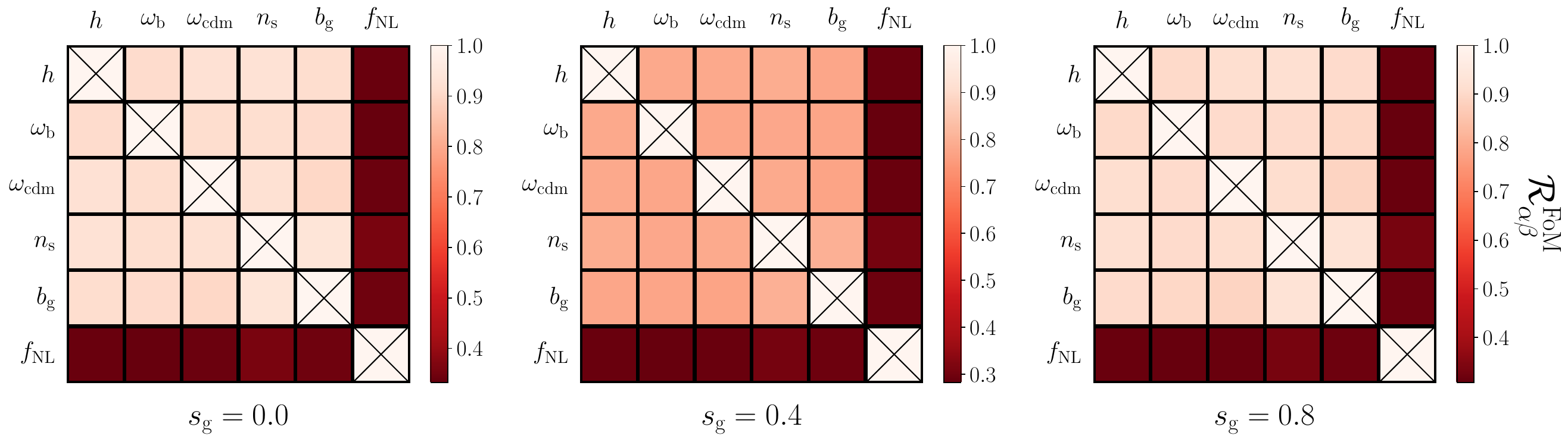}}
\caption{Ratio of the Figures-of-Merit computed without using and using the Limber approximation for the Euclid-like galaxy population with non-overlapping redshift bins. We show the cases $s_\mathrm{g}=0.0$ (\textit{left panel}), $s_\mathrm{g}=0.4$ (\textit{central panel}), $s_\mathrm{g}=0.8$ (\textit{right panel}). In all the three cases the maximum multipole is~$\ell_\mathrm{max}=3000$. Elements along the diagonal should be neglected since Figures-of-Merit are not well defined quantities for them.} 
\label{fig:ratio_fom_limber}
\end{figure}

For our straw-man survey the Limber approximation affects the magnitude of the Fisher matrix elements (some elements become larger, others smaller, others change sign). The overall effect on cosmological parameters is to change both the size and the correlation of the errors. In figure~\ref{fig:confidence_ellipses_limber} we show the $68\%$ CL marginalised constraints for all pairs of parameters for the Euclid-like galaxy population with non-overlapping redshift bins. Results are reported for two different choices of maximum multipole, $\ell_\mathrm{max}=180$ and~$\ell_\mathrm{max}=3000$ (following the methodology of \S~\ref{subsec:sky_maps_analysis}), in the $s_\mathrm{g}=0.6$ case. We find that these findings are not restricted to some particular choice of $s_\mathrm{g}$ and that results obtained from the uniform galaxy population are similar to those presented in figure~\ref{fig:confidence_ellipses_limber}. As in \S~\ref{subsec:cosmic_magnification}, the effects are more pronounced for the lower~$\ell_\mathrm{max}$ case.

\begin{figure}
\centerline{
\includegraphics[width=1.0\columnwidth]{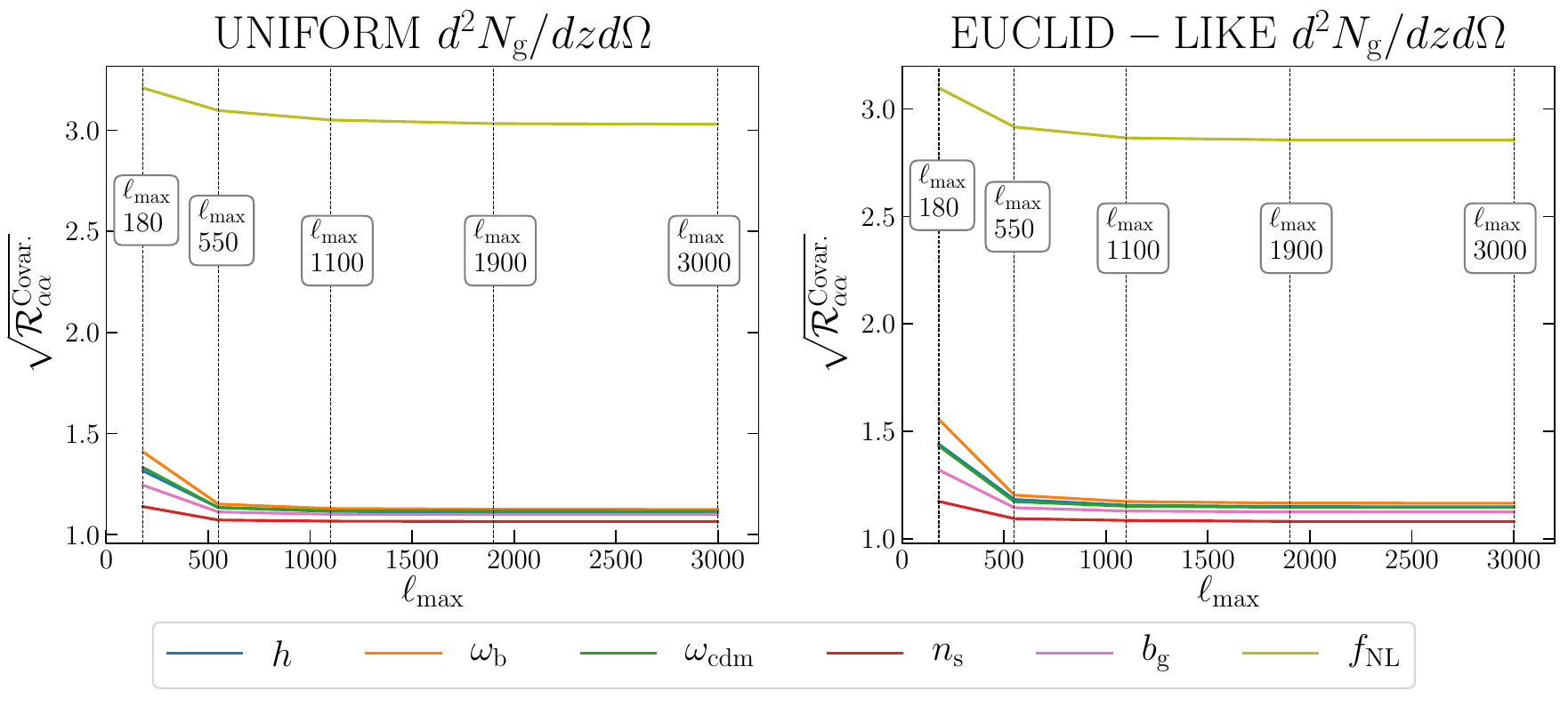}}
\caption{Evolution of the marginalized error ratio $\sqrt{\mathcal{R}^\mathrm{Covar.}_{\alpha\alpha}}$ as a function of the maximum multipole range chosen in the analysis for a uniform and an Euclid-like galaxy population with non-overlapping redshift bins in the \textit{left} and \textit{right panels}, respectively, when $s_\mathrm{g}=0.6$.} 
\label{fig:sigma_ratio_nolimber_vs_limber}
\end{figure}

It is interesting to note that using the Limber approximation always returns tighter constraints than with the exact computation of the integral. This effect becomes even more striking when looking at Figures-of-Merit, which ratio can be found in figure~\ref{fig:ratio_fom_limber}. For the Euclid-like galaxy population the Limber approximation overestimates the Figures-of-Merit, in particular those involving the~$f_\mathrm{NL}$ parameter are overestimated by~$30-40\%$, almost independently from the specific value of the magnification bias (we show only the $s_\mathrm{g}=\left\lbrace 0.0, 0.4, 0.8 \right\rbrace$ cases for practical purposes). Similar conclusions hold also for the uniform galaxy population case. The mis-estimation of errors does not disappear by increasing the maximum multipole included in the analysis, as we show in figure~\ref{fig:sigma_ratio_nolimber_vs_limber}. Similarly to the cosmic magnification case, we observe a ``saturation'' when all multipoles are included. Even if for standard cosmological parameters the systematic effect produced by the Limber approximation is of order~$10\%$, in the case of the~$f_\mathrm{NL}$ parameter we notice that we underestimate the real error by a factor~$3$ even for~$\ell_\mathrm{max}=3000$. The Limber approximation also affects degeneracies between different parameters by mis-estimating their degree of correlation, i.e., the confidence region orientation in the parameter space (see figure~\ref{fig:confidence_ellipses_limber}). We find that this effect is present for both galaxy populations and for different values of the magnification bias. 

%%%%%%%%%%%%%%%%%%%%%%%%%%%%%%%%%%%%%%%%%%%%%%%%%%%%%%%%%%%%%%%%%%%

\section{Effects of approximations in the multi-tracer analysis}
\label{sec:multitracing}

The framework introduced in section~\ref{sec:observable_modeling_fisher_matrix} is flexible enough to describe data coming from different surveys or tracers, provided that one can build an angular power spectrum (which is a standardized procedure given a  map).

\begin{figure}
\centerline{
\includegraphics[width=1.0\columnwidth]{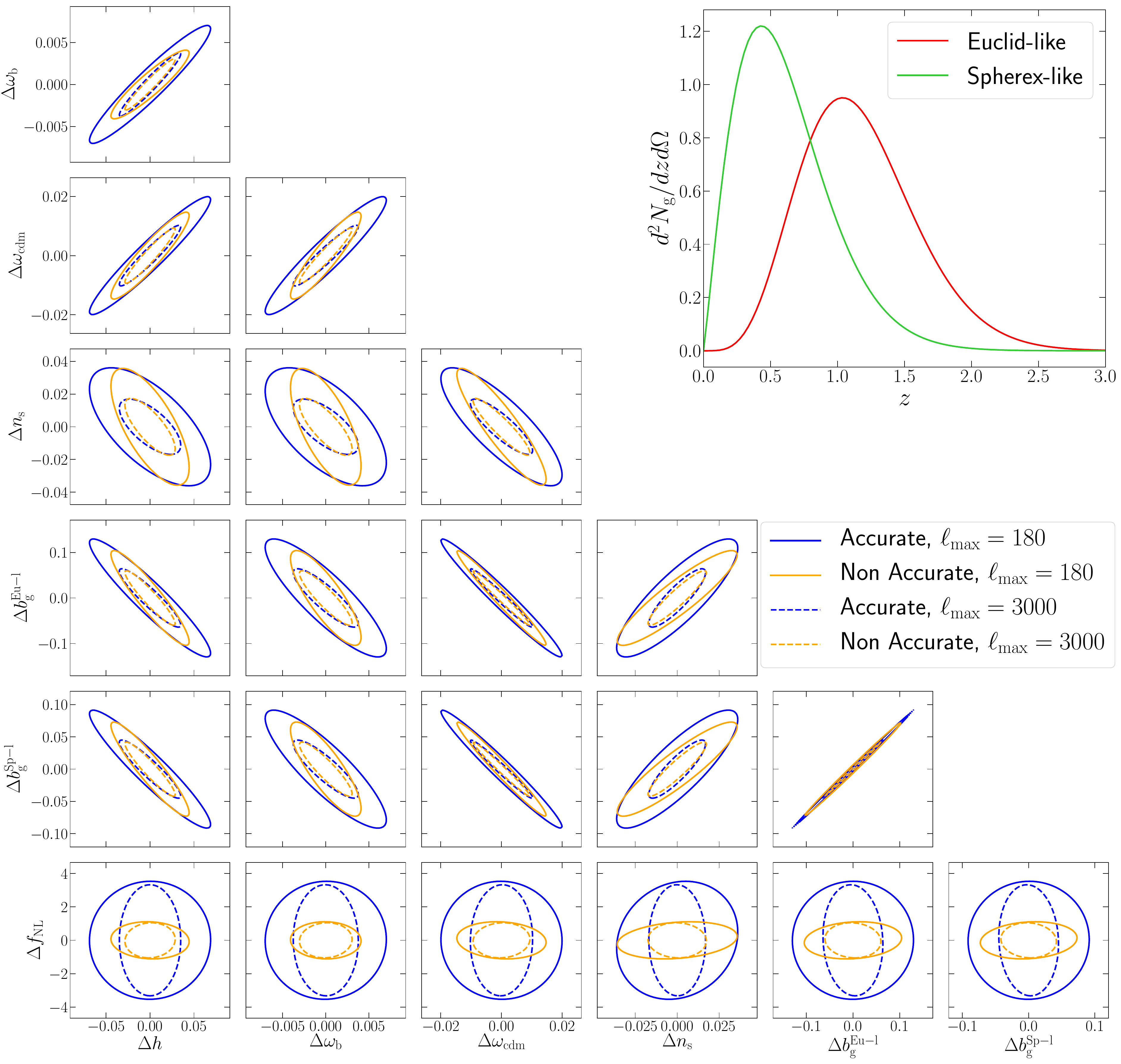}}
\caption{\textit{Upper right panel}: normalized galaxy redshift distribution~$d^2N_\mathrm{g}/dzd\Omega$ for the two surveys used in the multi-tracer analysis. \textit{Central triangle plot:} Two-dimensional marginalised $68\%$ CL regions for the combination of Euclid-like and SPHEREx-like galaxy surveys with non-overlapping redshift bins for two different maximum multipoles, $\ell_\mathrm{max}=180$ and~$\ell_\mathrm{max}=3000$. Orange solid and dashed lines refer to the non-accurate analysis in which we neglect off-diagonal terms of the covariance matrix, cosmic magnification and we used the Limber approximation, whereas blue solid and dashed lines refer to the accurate case, where none of these approximations are taken. This set of approximations yields an underestimation of the errors and in certain cases it slightly changes the parameter degeneracies.} 
\label{fig:confidence_ellipses_multitracing}
\end{figure}

We extended the public code \texttt{CLASS} to include the possibility of having different tracers, each one of them characterized by a different redshift distribution ($d^2N_X/dzd\Omega$) and by different bias parameters ($b_X,\ s_X,\ f^\mathrm{evo}_X$). This extension, called \texttt{Multi\_CLASS}, allows the user to specify these options, along with the standard ones connected with the choice of redshift binning and window functions, see e.g., appendix~\ref{app:multi_class}. Moreover, it can be also used to account for resolved gravitational wave events, as done in ref.~\cite{scelfo:gwxlss}.

We consider the combined analysis of an Euclid-like and a SPHEREx-like galaxy survey. The galaxy redshift distribution of these two surveys peaks at $z\lesssim 1$ for the SPHEREx-like and at $z\gtrsim 1$ for the Euclid-like, as can be seen in the upper right panel of figure~\ref{fig:confidence_ellipses_multitracing}. Therefore combining these experiments proves to be useful since it gives us access to the late-Universe structure formation history over a wide range of redshift. In this example we use non-overlapping redshift bins and we choose as magnification bias parameters the values $s^\mathrm{Eu-l}_\mathrm{g}=s^\mathrm{Sp-l}_\mathrm{g}=0.6$  for both the Euclid- and SPHEREx-like surveys. 

We compare two cases: one where none of the approximations presented in sections~\ref{sec:likelihood_modeling} and~\ref{sec:signal_modeling} is taken  and  one where all of them are taken at the same time (i.e., using a diagonal covariance matrix, neglecting lensing effects and using the Limber approximation). In figure~\ref{fig:confidence_ellipses_multitracing} we show the two-dimensional~$68\%$ CL regions  for two different maximum multipoles, $\ell_\mathrm{max}=180$ and~$\ell_\mathrm{max}=3000$, following the method of \S~\ref{subsec:sky_maps_analysis}. On can appreciate  that,  also in the multiple-tracers case, this set of approximations changes both the shape and the curvature of the likelihood around its maximum. As a result, the errors are mis-estimated, as well as the parameter degeneracies. In particular, this set of approximations underestimates the true statistical errors by a factor~$3$ for the~$f_\mathrm{NL}$ parameter. 

%%%%%%%%%%%%%%%%%%%%%%%%%%%%%%%%%%%%%%%%%%%%%%%%%%%%%%%%%%%%%%%%%%%

\section{Conclusions}
\label{sec:conclusions}

Forthcoming experiments promise to bring about new possibilities in the next decade. They aim not only to improve constraints on cosmological parameters within the standard model, but also to find signatures of new physics. Given the unprecedented sensitivity of on-going and forthcoming surveys, it is of fundamental importance to assess correctly the magnitude of statistical and systematic errors. In particular, obtaining unbiased estimates of cosmological parameters and their uncertainties is required in order to claim the detection of new physical effects.

In this paper we studied how common assumptions in the modelling of the likelihood and in the computation of the observable affect  the estimation of the curvature of the likelihood around its maximum and thus the estimated statistical errors on cosmological parameters. As a consequence, such approximations might invalidate not only Fisher matrix-based estimates (as demonstrated here) but also the actual parameter inference from real data. We showed that such approximations can change both the degeneracies between parameters and the size of the errors. In particular, we summarise our findings on the change of errors in tables~\ref{tab:results_section4}, \ref{tab:results_section5} and~\ref{tab:results_section6}, where we report the ratio of marginalised errors in the different cases we analysed. We notice that the error on the~$f_\mathrm{NL}$ parameter is particularly sensitive to many of these approximations. This is expected, since most of the approximations considered are inaccurate at large scales, where the contribution from primordial non-Gaussianity (parametrised by $f_{\rm NL}$) is the largest.

\begin{table}[h]
\centerline{
\begin{tabular}{|c|c|c|c|c|c||c|c|c|c|c|c|}
\multicolumn{12}{c}{NEGLECTING COVARIANCE - UNIFORM $d^2N_\mathrm{g}/dzd\Omega$} \\
\hline
\hline
\multicolumn{6}{c}{Overlapping Bins}              & \multicolumn{6}{c}{Non-Overlapping Bins} \\
\hline
\hline
$s_\mathrm{g}$ & 0.0 & 0.2 & 0.4 & 0.6 & 0.8 & $s_\mathrm{g}$ & 0.0 & 0.2 & 0.4 & 0.6 & 0.8 \\
\hline
$\sqrt{\mathcal{R}^\mathrm{Covar.}_{hh}}$ & 1.08 & 1.09 & 1.05 & 1.11 & 1.07 & $\sqrt{\mathcal{R}^\mathrm{Covar.}_{hh}}$ & 0.99 & 0.99 & 1.00 & 1.02 & 1.02 \\
$\sqrt{\mathcal{R}^\mathrm{Covar.}_{\omega_\mathrm{b}\omega_\mathrm{b}}}$ & 1.08 & 1.10 & 1.06 & 1.12 & 1.09 & $\sqrt{\mathcal{R}^\mathrm{Covar.}_{\omega_\mathrm{b}\omega_\mathrm{b}}}$ & 0.98 & 0.99 & 1.00 & 1.02 & 1.02 \\
$\sqrt{\mathcal{R}^\mathrm{Covar.}_{\omega_\mathrm{cdm}\omega_\mathrm{cdm}}}$ & 1.10 & 1.11 & 1.06 & 1.15 & 1.13 & $\sqrt{\mathcal{R}^\mathrm{Covar.}_{\omega_\mathrm{cdm}\omega_\mathrm{cdm}}}$ & 0.97 & 0.98 & 1.00 & 1.02 & 1.05 \\
$\sqrt{\mathcal{R}^\mathrm{Covar.}_{n_\mathrm{s} n_\mathrm{s}}}$ & 1.11 & 1.12 & 1.06 & 1.19 & 1.25 & $\sqrt{\mathcal{R}^\mathrm{Covar.}_{n_\mathrm{s} n_\mathrm{s}}}$ & 0.95 & 0.97 & 1.00 & 1.03 & 1.08 \\
$\sqrt{\mathcal{R}^\mathrm{Covar.}_{b_\mathrm{g} b_\mathrm{g}}}$ & 1.09 & 1.12 & 1.06 & 1.14 & 1.10 & $\sqrt{\mathcal{R}^\mathrm{Covar.}_{b_\mathrm{g} b_\mathrm{g}}}$ & 0.96 & 0.98 & 1.00 & 1.02 & 1.04 \\
$\sqrt{\mathcal{R}^\mathrm{Covar.}_{f_\mathrm{NL} f_\mathrm{NL}}}$ & 1.02 & 1.06 & 1.08 & 1.10 & 1.10 & $\sqrt{\mathcal{R}^\mathrm{Covar.}_{f_\mathrm{NL} f_\mathrm{NL}}}$ & 0.72 & 0.74 & 0.77 & 0.79 & 0.81 \\
\hline
\end{tabular}}
\vspace{0.5cm}
\centerline{
\begin{tabular}{|c|c|c|c|c|c||c|c|c|c|c|c|}
\multicolumn{12}{c}{NEGLECTING COVARIANCE - EUCLID-LIKE $d^2N_\mathrm{g}/dzd\Omega$} \\
\hline
\hline
\multicolumn{6}{c}{Overlapping Bins}              & \multicolumn{6}{c}{Non-Overlapping Bins} \\
\hline
\hline
$s_\mathrm{g}$ & 0.0 & 0.2 & 0.4 & 0.6 & 0.8 & $s_\mathrm{g}$ & 0.0 & 0.2 & 0.4 & 0.6 & 0.8 \\
\hline
$\sqrt{\mathcal{R}^\mathrm{Covar.}_{hh}}$ & 1.10 & 1.10 & 1.07 & 1.11 & 1.10 & $\sqrt{\mathcal{R}^\mathrm{Covar.}_{hh}}$ & 0.99 & 0.99 & 1.00 & 1.02 & 1.02 \\
$\sqrt{\mathcal{R}^\mathrm{Covar.}_{\omega_\mathrm{b}\omega_\mathrm{b}}}$ & 1.10 & 1.11 & 1.07 & 1.12 & 1.11 & $\sqrt{\mathcal{R}^\mathrm{Covar.}_{\omega_\mathrm{b}\omega_\mathrm{b}}}$ & 0.99 & 0.99 & 1.00 & 1.02 & 1.02 \\
$\sqrt{\mathcal{R}^\mathrm{Covar.}_{\omega_\mathrm{cdm}\omega_\mathrm{cdm}}}$ & 1.12 & 1.12 & 1.08 & 1.15 & 1.15 & $\sqrt{\mathcal{R}^\mathrm{Covar.}_{\omega_\mathrm{cdm}\omega_\mathrm{cdm}}}$ & 0.98 & 0.99 & 1.00 & 1.02 & 1.04 \\
$\sqrt{\mathcal{R}^\mathrm{Covar.}_{n_\mathrm{s} n_\mathrm{s}}}$ & 1.12 & 1.12 & 1.09 & 1.19 & 1.25 & $\sqrt{\mathcal{R}^\mathrm{Covar.}_{n_\mathrm{s} n_\mathrm{s}}}$ & 0.97 & 0.99 & 1.00 & 1.03 & 1.07 \\
$\sqrt{\mathcal{R}^\mathrm{Covar.}_{b_\mathrm{g} b_\mathrm{g}}}$ & 1.12 & 1.13 & 1.09 & 1.14 & 1.13 & $\sqrt{\mathcal{R}^\mathrm{Covar.}_{b_\mathrm{g} b_\mathrm{g}}}$ & 0.97 & 0.99 & 1.00 & 1.02 & 1.04 \\
$\sqrt{\mathcal{R}^\mathrm{Covar.}_{f_\mathrm{NL} f_\mathrm{NL}}}$ & 1.13 & 1.13 & 1.14 & 1.16 & 1.19 & $\sqrt{\mathcal{R}^\mathrm{Covar.}_{f_\mathrm{NL} f_\mathrm{NL}}}$ & 0.72 & 0.71 & 0.73 & 0.76 & 0.81 \\
\hline
\end{tabular}}
\caption{Summary of the results of section~\ref{sec:likelihood_modeling}. Ratio of marginalised errors computed including and neglecting the off-diagonal elements of the covariance matrix. Results are showed for a uniform (\textit{upper table}) and Euclid-like (\textit{lower table}) galaxy distribution with overlapping or non-overlapping redshift bins, assuming maximum multipole~$\ell_\mathrm{max}=3000$.}
\label{tab:results_section4}
\end{table}

\begin{table}[h]
\centerline{
\begin{tabular}{|c|c|c|c|c|c||c|c|c|c|c|c|}
\multicolumn{12}{c}{NEGLECTING COSMIC MAGNIFICATION} \\
\hline
\hline
\multicolumn{6}{c}{Uniform $d^2N_\mathrm{g}/dzd\Omega$} & \multicolumn{6}{c}{Euclid-like $d^2N_\mathrm{g}/dzd\Omega$} \\
\hline
\hline
$s_\mathrm{g}$ & 0.0 & 0.2 & 0.4 & 0.6 & 0.8 & $s_\mathrm{g}$ & 0.0 & 0.2 & 0.4 & 0.6 & 0.8 \\
\hline
$\sqrt{\mathcal{R}^\mathrm{Covar.}_{hh}}$ & 0.55 & 0.70 & 1.00 & 0.71 & 0.55 & $\sqrt{\mathcal{R}^\mathrm{Covar.}_{hh}}$ & 0.58 & 0.74 & 1.00 & 0.75 & 0.58 \\
$\sqrt{\mathcal{R}^\mathrm{Covar.}_{\omega_\mathrm{b}\omega_\mathrm{b}}}$ & 0.53 & 0.69 & 1.00 & 0.69 & 0.53 & $\sqrt{\mathcal{R}^\mathrm{Covar.}_{\omega_\mathrm{b}\omega_\mathrm{b}}}$ & 0.57 & 0.73 & 1.00 & 0.74 & 0.56 \\
$\sqrt{\mathcal{R}^\mathrm{Covar.}_{\omega_\mathrm{cdm}\omega_\mathrm{cdm}}}$ & 0.41 & 0.62 & 1.00 & 0.63 & 0.41 & $\sqrt{\mathcal{R}^\mathrm{Covar.}_{\omega_\mathrm{cdm}\omega_\mathrm{cdm}}}$ & 0.47 & 0.68 & 1.00 & 0.69 & 0.44 \\
$\sqrt{\mathcal{R}^\mathrm{Covar.}_{n_\mathrm{s} n_\mathrm{s}}}$ & 0.37 & 0.60 & 1.00 & 0.60 & 0.36 & $\sqrt{\mathcal{R}^\mathrm{Covar.}_{n_\mathrm{s} n_\mathrm{s}}}$ & 0.43 & 0.66 & 1.00 & 0.67 & 0.41 \\
$\sqrt{\mathcal{R}^\mathrm{Covar.}_{b_\mathrm{g} b_\mathrm{g}}}$ & 0.41 & 0.62 & 1.00 & 0.63 & 0.41 & $\sqrt{\mathcal{R}^\mathrm{Covar.}_{b_\mathrm{g} b_\mathrm{g}}}$ & 0.46 & 0.68 & 1.00 & 0.69 & 0.44 \\
$\sqrt{\mathcal{R}^\mathrm{Covar.}_{f_\mathrm{NL} f_\mathrm{NL}}}$ & 1.00 & 0.99 & 1.00 & 1.02 & 1.05 & $\sqrt{\mathcal{R}^\mathrm{Covar.}_{f_\mathrm{NL} f_\mathrm{NL}}}$ & 1.04 & 1.00 & 1.00 & 1.04 & 1.11 \\
\hline
\end{tabular}}
\vspace{0.5cm}
\centerline{
\begin{tabular}{|c|c|c|c|c|c||c|c|c|c|c|c|}
\multicolumn{12}{c}{USING LIMBER APPROXIMATION} \\
\hline
\hline
\multicolumn{6}{c}{Uniform $d^2N_\mathrm{g}/dzd\Omega$} & \multicolumn{6}{c}{Euclid-like $d^2N_\mathrm{g}/dzd\Omega$} \\
\hline
\hline
$s_\mathrm{g}$ & 0.0 & 0.2 & 0.4 & 0.6 & 0.8 & $s_\mathrm{g}$ & 0.0 & 0.2 & 0.4 & 0.6 & 0.8 \\
\hline
$\sqrt{\mathcal{R}^\mathrm{Covar.}_{hh}}$ & 1.05 & 1.10 & 1.23 & 1.11 & 1.06 & $\sqrt{\mathcal{R}^\mathrm{Covar.}_{hh}}$ & 1.08 & 1.13 & 1.25 & 1.15 & 1.09 \\
$\sqrt{\mathcal{R}^\mathrm{Covar.}_{\omega_\mathrm{b}\omega_\mathrm{b}}}$ & 1.06 & 1.11 & 1.25 & 1.12 & 1.07 & $\sqrt{\mathcal{R}^\mathrm{Covar.}_{\omega_\mathrm{b}\omega_\mathrm{b}}}$ & 1.09 & 1.15 & 1.27 & 1.16 & 1.10 \\
$\sqrt{\mathcal{R}^\mathrm{Covar.}_{\omega_\mathrm{cdm}\omega_\mathrm{cdm}}}$ & 1.05 & 1.10 & 1.24 & 1.12 & 1.06 & $\sqrt{\mathcal{R}^\mathrm{Covar.}_{\omega_\mathrm{cdm}\omega_\mathrm{cdm}}}$ & 1.07 & 1.13 & 1.25 & 1.15 & 1.08 \\
$\sqrt{\mathcal{R}^\mathrm{Covar.}_{n_\mathrm{s} n_\mathrm{s}}}$ & 1.00 & 1.05 & 1.19 & 1.07 & 1.02 & $\sqrt{\mathcal{R}^\mathrm{Covar.}_{n_\mathrm{s} n_\mathrm{s}}}$ & 1.01 & 1.06 & 1.18 & 1.08 & 1.02 \\
$\sqrt{\mathcal{R}^\mathrm{Covar.}_{b_\mathrm{g} b_\mathrm{g}}}$ & 1.03 & 1.08 & 1.23 & 1.10 & 1.05 & $\sqrt{\mathcal{R}^\mathrm{Covar.}_{b_\mathrm{g} b_\mathrm{g}}}$ & 1.05 & 1.11 & 1.23 & 1.12 & 1.06 \\
$\sqrt{\mathcal{R}^\mathrm{Covar.}_{f_\mathrm{NL} f_\mathrm{NL}}}$ & 2.91 & 2.94 & 2.98 & 3.03 & 3.08 & $\sqrt{\mathcal{R}^\mathrm{Covar.}_{f_\mathrm{NL} f_\mathrm{NL}}}$ & 2.76 & 2.73 & 2.78 & 2.85 & 2.97 \\
\hline
\end{tabular}}
\caption{Summary of the results of section~\ref{sec:signal_modeling}. Ratio of marginalised errors computed including and neglecting cosmic magnification (\textit{upper table}) and not using and using the Limber approximation (\textit{lower table}). Results are showed for a uniform and Euclid-like galaxy distribution with non-overlapping redshift bins, assuming maximum multipole~$\ell_\mathrm{max}=3000$.}
\label{tab:results_section5}
\end{table}

\begin{table}[h]
\centerline{
\begin{tabular}{|c|c|}
\multicolumn{2}{c}{MULTI-TRACER ANALYSIS} \\
\hline
\hline
\multicolumn{2}{c}{Euclid-like and SPHEREx-like $d^2N_\mathrm{g}/dzd\Omega$} \\
\hline
\hline
$\sqrt{\mathcal{R}^\mathrm{Covar.}_{hh}}$ & 1.22 \\
$\sqrt{\mathcal{R}^\mathrm{Covar.}_{\omega_\mathrm{b}\omega_\mathrm{b}}}$ & 1.24  \\
$\sqrt{\mathcal{R}^\mathrm{Covar.}_{\omega_\mathrm{cdm}\omega_\mathrm{cdm}}}$ & 1.10  \\
$\sqrt{\mathcal{R}^\mathrm{Covar.}_{n_\mathrm{s} n_\mathrm{s}}}$ & 0.97  \\
$\sqrt{\mathcal{R}^\mathrm{Covar.}_{b^\mathrm{Eu-l}_\mathrm{g} b^\mathrm{Eu-l}_\mathrm{g}}}$ & 1.07  \\
$\sqrt{\mathcal{R}^\mathrm{Covar.}_{b^\mathrm{Sp-l}_\mathrm{g} b^\mathrm{Sp-l}_\mathrm{g}}}$ & 1.07  \\
$\sqrt{\mathcal{R}^\mathrm{Covar.}_{f_\mathrm{NL} f_\mathrm{NL}}}$ & 3.16  \\
\hline
\end{tabular}}
\caption{Summary of the results of section~\ref{sec:multitracing}. Ratio of marginalised errors computed when taking none of the assumptions described in sections~\ref{sec:likelihood_modeling} and~\ref{sec:signal_modeling} and when adopting all of them. Results are showed for an Euclid-like and SPHEREx-like galaxy distribution with non-overlapping redshift bins, assuming maximum multipole~$\ell_\mathrm{max}=3000$.}
\label{tab:results_section6}
\end{table}

This paper does not aim to cover all possible sources of uncertainty. Future galaxy survey will probe also the quasi-linear and non-linear regimes. However, the modelling of these scales is not as robust as linear theory: one possibility to deal with these uncertainties is to include a theoretical error on the RHS of equation~\eqref{eq:total_angular_power_spectrum}. The authors of refs.~\cite{audren:theoreticalerrors, baldauf:theoreticalerrors, sprenger:theoreticalerrors} showed that the constraints obtained when non-linear scales are included degrade by a factor of a few when theoretical errors/uncertainties are included in the analysis. Hence, in this case, inaccurate modelling of the noise results in a mis-estimate of marginalized errors of the same magnitude of the cases discussed in this work.

In particular, modifying parameter degeneracies becomes very relevant when considering the complementarity between different data sets. For instance, it is customary to combine large-scale structure constraints with e.g., the cosmic microwave background ones to break parameter degeneracies. Mis-estimating these degeneracies invalidates the assessment of how they can be broken when different observables are combined.

We proved that the robustness of an approximation cannot be immediately judged by looking at the Fisher matrix, since small differences in many off-diagonal terms can add up to create a considerable effect on the marginalised error, as e.g., in the case of cosmic magnification. Moreover, the final estimates can be biased even when the adopted approximations involve effects that appear intuitively irrelevant or are independent on the parameters of interest. 

The changes of error size should not be seen only as a problem per se, but also in light of the second effect approximations have, i.e., the induced shift on the best-fit parameters. This is extensively covered in the companion paper~\cite{bernal:bestfitshift}, however we summarise here the main point. As the reader can appreciate from figure 6 in ref.~\cite{bernal:bestfitshift}, the typical shift in the best-fit parameters is of order of the 1~$\sigma$ error in the ``cosmic magnification'' and ``Limber approximation'' cases\footnote{Alternatively, in the language of ref.~\cite{bernal:bestfitshift}, the shift normalised with respect to the error is of order unity, i.e.~$\Delta_\mathrm{syst}/\sigma \simeq \mathcal{O}(1)$.}. Therefore, a change of order few in the size of the errors (as we show it is the case) is sufficient to shift the best-fit parameters two-three sigma away from their real value, creating what it might appear as a (artificial) ``tension'' between different datasets. The underlying risk is to interpreted as ``new physics'' what in reality is a systematic effect.

In this work we focused on future galaxy surveys and a simple extension to the standard~$\Lambda$CDM model, i.e., including local primordial non-Gaussianity, parametrized by~$f_\mathrm{NL}$. However, the main message of this work and its implications are far more general, especially in the case where small signatures of new physics are sought. This is particularly important not only in the already existing multi-tracer cosmology~\cite{blake:gamamultitracer, beutler:bossmultitracerI, marin:bossmultitracerII} but also in the newly emergent multi-messenger era of cosmology~\cite{abbott:multimessenger, aartsen:multimessenger}. The great statistical power reached when combination of different tracers, or of the same tracer detected using different ``messengers'', is accompanied by added complications in the modelling, hence it is tempting to take several approximations for simplicity and speed. Given the strong observational effort in building more powerful astrophysical and cosmological experiments in the next decade, it is of paramount importance to model correctly the target observables, as well as the likelihood used for cosmological inference. We stress that any modelling approximation should be thoroughly tested; the kind of analyses presented here should be performed for all observables and approximations, such as e.g., the flat-sky approximation in galaxy correlation functions, the effects of cosmological perturbations on gravitational waves propagation, and many more. 

As a result of our effort, and to make the above program easier to carry out, we are presenting and releasing the code \texttt{Multi\_CLASS}: the first Boltzmann code based on \texttt{CLASS} that allows the computation of cross angular power spectra of different tracers (or messengers). We envision it will be a useful tool for forecasts and real data analyses once future datasets become available.

%%%%%%%%%%%%%%%%%%%%%%%%%%%%%%%%%%%%%%%%%%%%%%%%%%%%%%%%%%%%%%%%%%%

\newpage

\begin{acknowledgments}
Funding for this work was partially provided by the Spanish MINECO under projects AYA2014-58747-P AEI/FEDER, UE, and MDM-2014-0369 of ICCUB (Unidad de Excelencia Mar\'ia de Maeztu). NB is supported by the Spanish MINECO under grant BES-2015-073372. JLB is supported by  the Allan C. and Dorothy H. Davis Fellowship, and has been supported by the Spanish MINECO under grant BES-2015-071307, co-funded by the ESF during part of the development of this work. AR has received funding from the People Programme (Marie Curie Actions) of the European Union H2020 Programme under REA grant agreement number 706896 (COSMOFLAGS). GS was supported by the Erasmus+ for Trainership grant during the early stages of this work, subsequently by grant from the ``Maria de Maeztu de Ci\`encies del Cosmos'' project mentioned above. GS is supported by the INFN INDARK PD51 grant. LV acknowledges support by European Union's Horizon 2020 research and innovation programme ERC (BePreSySe, grant agreement 725327).
\end{acknowledgments}

%%%%%%%%%%%%%%%%%%%%%%%%%%%%%%%%%%%%%%%%%%%%%%%%%%%%%%%%%%%%%%%%%%%

\appendix

\section{Relativistic Number Counts}
\label{app:relativistic_number_counts}
In this appendix we explicitly list the contributions to the galaxy number counts, following the notation of ref. \cite{didio:classgal}. The transfer functions of equation \eqref{eq:numbercount_fluctuation} read as 
\begin{equation}
\begin{aligned}
\Delta_\ell^\mathrm{den}(k,z) &= b_{X,\mathrm{tot}} \delta(k,\tau_z) j_\ell,	\\
\Delta_\ell^\mathrm{vel}(k,z) &= \Delta_\ell^\mathrm{rsd}(k,z) + \Delta_\ell^\mathrm{dop}(k,z),	\\
\Delta_\ell^\mathrm{rsd}(k,z) &=  \frac{k}{\mathcal{H}}\frac{d^2j_\ell}{dy^2}  V(k,\tau_z),	\\
\Delta_\ell^\mathrm{dop}(k,z) &= \left[(f^\mathrm{evo}_X-3)\frac{\mathcal{H}}{k}j_\ell + \left(\frac{\mathcal{H}'}{\mathcal{H}^2}+\frac{2-5s_X}{r(z)\mathcal{H}}+5s_X-f^\mathrm{evo}_X\right)\frac{dj_\ell}{dy} \right]  V(k,\tau_z),	\\
\Delta_\ell^\mathrm{len}(k,z) &= \ell(\ell+1) \frac{2-5s_X}{2} \int_0^{r(z)} dr \frac{r(z)-r}{r(z) r} \left[\Phi(k,\tau_z)+\Psi(k,\tau_z)\right] j_\ell(kr),	\\
\Delta_\ell^\mathrm{gr}(k,z)  &= \left[\left(\frac{\mathcal{H}'}{\mathcal{H}^2}+\frac{2-5s_X}{r(z)\mathcal{H}}+5s_X-f^\mathrm{evo}_X+1\right)\Psi(k,\tau_z) + \left(-2+5s_X\right) \Phi(k,\tau_z) + \mathcal{H}^{-1}\Phi'(k,\tau_z)\right] j_\ell + \\
&+ \int_0^{r(z)} dr \frac{2-5s_X}{r(z)} \left[\Phi(k,\tau)+\Psi(k,\tau)\right]j_\ell(kr) , \\
&+ \int_0^{r(z)} dr \left(\frac{\mathcal{H}'}{\mathcal{H}^2}+\frac{2-5s_X}{r(z)\mathcal{H}}+5s_X-f^\mathrm{evo}_X\right)_{r(z)} \left[\Phi'(k,\tau)+\Psi'(k,\tau)\right] j_\ell(kr).
\end{aligned}
\end{equation}
According to the notation of ref. \cite{didio:classgal}, $r$ is the conformal distance, $\tau=\tau_0-r$ is the conformal time, $\tau_z=\tau_0-r(z)$, $b_{X,\mathrm{tot}}$ is the total bias parameter, $s_X$ is the magnification bias parameter, $f^\mathrm{evo}_X$ is the evolution bias parameter, Bessel functions and their derivatives $j_\ell$, $\frac{dj_\ell}{dy}$, $\frac{d^2j_\ell}{dy^2}$ are evaluated at $y=kr(z)$ unless explicitly stated, $\mathcal{H}$~is the conformal Hubble parameter, a prime~$'$ indicates derivatives with respect to conformal time, $\delta$ is the density contrast in comoving gauge, $V$ is the peculiar velocity, and $\Phi$ and $\Psi$ are the Bardeen potentials.
 
The velocity term $\Delta_\ell^\mathrm{vel}(k,z)$ has been written in terms of the pure (Kaiser) redshift-space distortions term $\Delta_\ell^\mathrm{rsd}(k,z)$ and in term of Doppler contributions $\Delta_\ell^\mathrm{dop}(k,z)$. The magnification and evolution bias parameters enter only in the Doppler term, whereas the Kaiser term does not depend on the parameter of the tracer.

%%%%%%%%%%%%%%%%%%%%%%%%%%%%%%%%%%%%%%%%%%%%%%%%%%%%%%%%%%%%%%%%%%%

\section{Description of \texttt{Multi\_CLASS}}
\label{app:multi_class}

In this appendix we explain how the \texttt{Multi\_CLASS}\footnote{The code will be publicly released after the article is accepted. Users can find and download the code on the GitHub page \url{https://github.com/nbellomo/Multi_CLASS}.} code is structured, available features in this initial release of the code and more technical details on the modifications introduced. \texttt{Multi\_CLASS} is based on \texttt{CLASS}, therefore it can be used as \texttt{CLASS} itself, unless otherwise stated.

\texttt{Multi\_CLASS} is the first public Boltzmann code that allows to compute the cross-tracer angular power spectrum for multiple galaxy (and other tracers) populations. The code allows the user to specify, for each tracer, its own number density redshift distribution, bias, magnification bias and evolution bias. Moreover, we implemented also the effect of primordial non-Gaussianity of the local-type, parametrised by~$f_\mathrm{NL}$, on the tracer bias.

%%%%%%%%%%%%%%%%%%%%%%%%%%%%%%%%%%%%%%%%%%%%%%%%%%%%%%%%%%%%%%%%%%%

\subsection{\texttt{Multi\_CLASS} for users}
We list here the input options that can be used in the \texttt{.ini} file and we report between squared parenthesis~\texttt{[...]} the input that can be introduced by the user. Some of the input parameter names have been changed with respect to the standard version of \texttt{CLASS}, to increase the transparency and to reflect the inner structure of the code.

The spirit of \texttt{Multi\_CLASS} is that \textit{``all the input parameters should be explicitly declared''}. Some options available in \texttt{CLASS} have been removed, e.g., the possibility to declare the bin width only once, relying on the code to assign it to all the redshift bins. The ultimate goal is to avoid any possible ambiguity, even if the level of conciseness in the \texttt{.ini} file has been reduced.

To access \texttt{Multi\_CLASS} options, the user must require for the number count angular power spectra to be computed (\texttt{output = nCl}) and must declare which physical effects should be included in the $C_\ell$ (\texttt{number count contributions = density, rsd, lensing, gr}), \textit{cf.} equation~\eqref{eq:numbercount_fluctuation} and appendix~\ref{app:relativistic_number_counts}. The parameters the user can specify are listed below.

\begin{enumerate}
\item \texttt{selection\_multitracing}: fix the number of tracers~$N_\mathrm{tracers}$ considered. The option \texttt{[yes]} allows for two different tracers ($N_\mathrm{tracers}=2$), whereas \texttt{[no]} runs the code with the standard single-tracer method ($N_\mathrm{tracers}=1$). The default value is \texttt{[no]}, i.e., single-tracer.

\item \texttt{selection\_mean}: list of the mean redshift~$z_i$ for different redshift bins, \textit{cf.} equation~\eqref{eq:Delta_l}. As in \texttt{CLASS}, the user must provide a \texttt{[list of numbers]} separated by a comma. The length of the list sets the number of redshift bins~$N_\mathrm{bins}$ considered. There is no default value, so a number or a list of numbers must be specified.

\item \texttt{selection\_width}: list of redshift bin widths~$\Delta z$, \textit{cf.} equation~\eqref{eq:Delta_l}. Contrary to \texttt{CLASS}, the user must provide a \texttt{[list of numbers]}, separated by a comma, of the same length of the list given in \texttt{selection\_mean}. No default value is assigned.

\item \texttt{selection\_window}: window function~$W(z,z_i,\Delta_z)$ used for both tracers, \textit{cf.} equation~\eqref{eq:Delta_l}. As in \texttt{CLASS}, the options available are \texttt{[gaussian, tophat, dirac]}. There is no default value, so a window function must be provided. The name of the corresponding parameter in \texttt{CLASS} was simply \texttt{selection}.

\item \texttt{selection\_bias}: list of mean bias~$b_X$ parameter in different redshift bins, \textit{cf.} equation~\eqref{eq:total_galaxy_bias}. The user must provide a \texttt{[list of numbers]}, separated by a comma, containing the value of the bias parameter in each redshift bin. There is no default value, and the length of the list must be $N_\mathrm{bins} \times N_\mathrm{tracers}$. In the case of two tracers, the bias values should be ordered as
\begin{equation}
\left[b_X(z_1), \cdots, b_X(z_N), b_Y(z_1), \cdots, b_Y(z_N) \right].
\end{equation}

\item \texttt{selection\_magnification\_bias}: list of mean magnification bias~$s_X$ parameter in different redshift bins, \textit{cf.} equation~\eqref{eq:galaxymagnificationbias}. The user must provide a \texttt{[list of numbers]}, separated by a comma, containing the value of the magnification bias parameter in each redshift bin. There is no default value, and the length of the list must be $N_\mathrm{bins} \times N_\mathrm{tracers}$. In the case of two tracers, the magnification bias values should be ordered as
\begin{equation}
\left[s_X(z_1), \cdots, s_X(z_N), s_Y(z_1), \cdots, s_Y(z_N) \right].
\end{equation}

\item \texttt{selection\_dNdz\_1} and \texttt{selection\_dNdz\_2}: select the source number density~$dN_X/dz$ per redshift bin, \textit{cf.} equation~\eqref{eq:Delta_l}. Choose the \texttt{[analytic]} input to select between one of the hardcoded number densities (see point 8), or the \texttt{[file]} input to read the distribution from a file (see point 9 and 10). The code always reads the \texttt{selection\_dNdz\_1} option, whereas the \texttt{selection\_dNdz\_2} input is read only if \texttt{selection\_multitracing = yes}. If \texttt{dNdz\_selection\_1} is left unspecified in the single-tracer case, the code uses a uniform~$dN_X/dz$. In the multi-tracer case the user must always specify the \texttt{[analytic/file]} option for both tracers, and it must be the same for both of them\footnote{This particular restriction will be removed in future releases of the code.}. The input option \texttt{dNdz\_selection} is not supported any more, in order to decrease the number of different inputs. In this way the user can switch from multiple tracers to a single one just by using the \texttt{selection\_multitracing} option.

\item \texttt{selection\_tracer\_1} and \texttt{selection\_tracer\_2}: select between the included catalog of hardcoded tracer number density redshift distributions, in case of selecting \texttt{selection\_dNdz\_1 = analytic} and/or \texttt{selection\_dNdz\_2 = analytic}. If \texttt{selection\_multitracing = yes}, the code will also read \texttt{selection\_tracer\_2}. The catalog includes the number density redshift distributions corresponding to \texttt{[euclid\_galaxy]}~\cite{blanchard:euclidforecasts} and \texttt{[spherex\_galaxy]}~\cite{dore:spherexwhitepaperI}. It also includes the number density redshift distribution of gravitational waves generated by astrophysical sources, \texttt{[astrophysical\_gws]}~\cite{scelfo:gwxlss}, as explain in \S~\ref{subapp:gws}. New redshift distributions can be consistently hardcoded in the \texttt{transfer\_dNdz\_analytic} and \texttt{transfer\_dln\_dNdz\_dz\_analytic} functions, both contained in the \texttt{transfer.c} module.

\item \texttt{selection\_dNdz\_filepath\_1} and \texttt{selection\_dNdz\_filepath\_2}: path to the file containing the source number density per redshift bin. The file should contain two columns, $(z, dN_X/dz)$, as in standard \texttt{CLASS}. The path to the second file is read only if\\ \texttt{selection\_multitracing = yes}.

\item \texttt{selection\_dNdzevolution\_filepath\_1} and \texttt{selection\_dNdzevolution\_filepath\_2}: path to the file containing the source number density per redshift bin used to compute the evolution bias parameter. The file should contain two columns, $(z, dN_X/dz)$, as in standard \texttt{CLASS}. The second path is read only if \texttt{selection\_multitracing = yes}.

\item \texttt{non\_diagonal}: similarly to \texttt{CLASS}, it allows for the computation of the cross-bin angular power spectrum, i.e.,~$C^{XY}_\ell(z_i,z_j)$, with~$i \neq j$ and $X\neq Y$ ($X=Y$) if \texttt{selection\_multitracing = yes} (\texttt{selection\_multitracing = no}). The default value is \texttt{[0]}.
\end{enumerate}

\textbf{Remark \#1}: the \texttt{dNdz\_evolution} option has been removed. The user can specify the evolution bias~$f^\mathrm{evo}_X$ either by hardcoding its value in the \texttt{transfer\_dln\_dNdz\_dz\_analytic} function or by providing a file with the observed number density. 

\textbf{Remark \#2}: the \texttt{.ini} file used to compute the cross-tracer angular power spectra can automatically be used to compute the single tracer angular power spectrum for the \underline{first} tracer by switching off the multitracing option.

%%%%%%%%%%%%%%%%%%%%%%%%%%%%%%%%%%%%%%%%%%%%%%%%%%%%%%%%%%%%%%%%%%%

\subsection{Primordial Non-Gaussianity}

\texttt{Multi\_CLASS} includes also the possibility to compute the contribution to tracer bias of non-Gaussianities of the local type. This kind of non-Gaussianity is a natural prediction of the simplest single-field slow-roll inflationary models~\cite{acquaviva:ngfrominflation, maldacena:ngfrominflation}. It is described in real space by the well-known quadratic model~\cite{gangui:ngfrominflation, verde:ngfrominflation, komatsu:ngfrominflation} as second-order non-Gaussian corrections of the total gravitational potential~$\phi_\mathrm{NG}$ and they are typically parametrized by $f_\mathrm{NL}$. There are two conventions widely used in the literature:
\begin{equation}
\begin{aligned}
\phi_{\mathrm{NG},p} &= \phi_p + f^\mathrm{p}_\mathrm{NL} \left(\phi^2_p + \left\langle \phi^2_p \right\rangle \right), \\
\phi_{\mathrm{NG},0} &= \phi_0 + f^\mathrm{LSS}_\mathrm{NL} \left(\phi^2_0 + \left\langle \phi^2_0 \right\rangle \right).
\end{aligned}
\end{equation}
In the first one the expansion is done in terms of the primordial Gaussian gravitational potential~$\phi_p$, whereas in the second one it is done in terms of the primordial Gaussian gravitational potential linearly extrapolated at redshift~$z=0$, i.e., $\phi_0 = D(z=0)\phi_p$, where in this case the linear growth factor has not been normalized to unity at redshift~$z=0$. The $f_\mathrm{NL}$ parameters of the two expansions are connected by $f^\mathrm{p}_\mathrm{NL} = D(z=0) \times f^\mathrm{LSS}_\mathrm{NL}$.

Following the approach of ref.~\cite{dalal:ngbias}, the overdensity can be written in Lagrangian space as
\begin{equation}
\delta_\mathrm{g} = b_L\left(1 + 2 f^\mathrm{p}_\mathrm{NL} \delta_\mathrm{crit} S^{-1}_{\delta_\mathrm{m}} \right) \delta_\mathrm{m},
\end{equation}
where $b_L$ is the Lagrangian linear bias, $\delta_\mathrm{crit}$ is the linearly extrapolated overdensity for gravitational collapse ($\delta_\mathrm{crit}=1.686$ for spherical collapse in an Einstein-de Sitter cosmology) and $S_{\delta_\mathrm{m}} = \delta_\mathrm{m}/\phi_p$ is called ``matter source function'', following \texttt{CLASS} nomenclature. Therefore, without introducing any approximation on the transfer function, the Eulerian bias implemented in \texttt{Multi\_CLASS} is given by 
\begin{equation}
b_\mathrm{E, tot} = b_\mathrm{E} + 2(b_\mathrm{E}-1)f^\mathrm{LSS}_\mathrm{NL}D(z=0)\delta_\mathrm{crit} S^{-1}_{\delta_\mathrm{m}},
\end{equation}
which reduces to equation~\eqref{eq:total_galaxy_bias} when $D(z=0) S^{-1}_{\delta_\mathrm{m}}$ is written explicitly for an Einstein-de Sitter cosmology. According to the predictions suggested by some inflationary models~\cite{shandera:scaledependentbias, raccanelli:scaledependentbias}, an additional scale dependence to the non-linear parameter through the change~$f_\mathrm{NL}\to f_\mathrm{NL}\times(k/k_\mathrm{NG})^{n_\mathrm{NG}}$ has been proposed, where~$k_\mathrm{NG}$ is the non-Gaussianity pivot scale and~$n_\mathrm{NG}$ is the tilt or the running of non-Gaussianities.

The different parameters the user can include in the computation of the bias are

\begin{enumerate}
\item \texttt{f\_NL}: amplitude of non-Gaussian correction to the Newtonian gravitational potential at the pivot scale in the ``LSS'' convention, i.e., this parameter is $f^\mathrm{LSS}_\mathrm{NL}$. The default value is $f_\mathrm{NL} = 0.0$.

\item \texttt{n\_NG}: tilt of the non-Gaussian correction. The default value is $n_\mathrm{NG} = 0.0$.

\item \texttt{k\_pivot\_NG}: pivot scale of the non-Gaussian correction. The default value is $k_\mathrm{NG} = 1.0\ \mathrm{Mpc^{-1}}$.
\end{enumerate}

%%%%%%%%%%%%%%%%%%%%%%%%%%%%%%%%%%%%%%%%%%%%%%%%%%%%%%%%%%%%%%%%%%%

\subsection{Gravitational waves from astrophysical sources}
\label{subapp:gws}

\texttt{Multi\_CLASS} can use as tracer also the resolved gravitational wave (GW) events, as done in refs.~\cite{scelfo:gwxlss, calore:gwxlss}. We include in \texttt{Multi\_CLASS} also an additional tracer besides galaxies: GWs coming from astrophysical sources. These sources are most likely located in galaxies which undergo an intense star formation history, hence resolved GW events trace the large-scale structure of the Universe.

Following ref.~\cite{scelfo:gwxlss}, the GW number density redshift distribution can be written as
\begin{equation}
\frac{d^2N_\mathrm{GW}}{dzd\Omega} = T_\mathrm{obs} \frac{c\ r^2(z)}{(1+z)H(z)} \mathcal{R}_\mathrm{tot}(z) F^\mathrm{detectable}_\mathrm{GW}(z),
\label{eq:gw_number_density}
\end{equation}
where~$T_\mathrm{obs}$ is the total observation time, $c$ is the speed of light, $r(z)$ is the comoving distance, $H(z)$ is the Hubble expansion rate, $\mathcal{R}_\mathrm{tot}(z)$ is the total comoving merger rate and~$F^\mathrm{detectable}_\mathrm{GW}(z)$ is the fraction of detectable events, which depends on the specifics of GW observatory under consideration.

In the example introduced in the code we assume Einstein Telescope as GW detector, hence we assume that all the GW events up to redshift $z_\mathrm{max} \simeq 5$ are detected, i.e., $F^\mathrm{detectable}_\mathrm{GW}(z) \simeq \Theta(5-z)$, where $\Theta$ is the Heaviside Theta function. We choose as total merger rate the fiducial model of ref.~\cite{mapelli:mergerrate}. Assuming an expansion history close to the~$\Lambda$CDM one and~$T_\mathrm{obs}=1\ \mathrm{yr}$, the shape of equation~\eqref{eq:gw_number_density} can be approximated by the one reported in equation~\eqref{eq:tracer_distribution} with $\mathcal{A}=29700\ \mathrm{GWs / deg^2},\ z_0=2.55,\ \alpha=1.3,\ \beta=2.0$. The approximation is valid for~$z\lesssim 3$;  the user has to find a more refined approximation for higher redshiftn. However, we note that this approximation is certainly sufficient to cross-correlate GWs catalogs with SPHEREx/Euclid/DESI-like galaxy surveys.

The bias and magnification bias parameters for GWs depend on the considered scenario, whereas the evolution bias can be derived directly from equation~\eqref{eq:gw_number_density}. We refer the interested reader to ref.~\cite{scelfo:gwxlss}, where typical values of these parameters are discussed in details.

In the case of present and future GW detectors the maximum multipole is always~$\ell_\mathrm{max}\lesssim 100$, see e.g., ref.~\cite{scelfo:gwxlss} and refs. therein.

%%%%%%%%%%%%%%%%%%%%%%%%%%%%%%%%%%%%%%%%%%%%%%%%%%%%%%%%%%%%%%%%%%%

\subsection{\texttt{Multi\_CLASS} for developers}

In this section we present more technical details for developers and/or users that want to further modify the code.

\begin{enumerate}

\item The logic underlying the ordering of indices when filling arrays is \textit{``First everything related to the first tracer, then everything related to the second tracer''}.

\item The variables \texttt{selection\_num}, \texttt{selection\_mean}, \texttt{selection\_width}, \texttt{selection\_window}, \texttt{selection\_bias} and \texttt{selection\_magnification\_bias} are now defined in the \texttt{transfers} structure, since they are not needed by the \texttt{perturbs} structure. The only variable (related to the number count power spectrum) really needed in \texttt{perturbation.c} is the new variable \texttt{smallest\_redshift\_bin}, which contains the smallest mean redshift declared in \texttt{selection\_mean}.

\item The function \texttt{transfer\_dNdz\_analytic} has been split into two functions with two different tasks: \texttt{transfer\_dNdz\_analytic} and \texttt{transfer\_dln\_dNdz\_dz\_analytic}.

\item The list of input parameters of certain functions has been changed, for instance the \texttt{bin} input parameter has been removed from the \texttt{transfer\_source\_resample} function since it was not needed; or the \texttt{tracer} input has been added to many functions of the \texttt{transfer.c} module.

\item There are functions, e.g., \texttt{transfer\_selection\_times}, that need only the redshift \texttt{bin} input in the current version of \texttt{Multi\_CLASS}. In future versions, when there will be the possibility to have different redshift bins for different tracers, these function will likely need also the \texttt{tracer} input parameter.

\end{enumerate}

%%%%%%%%%%%%%%%%%%%%%%%%%%%%%%%%%%%%%%%%%%%%%%%%%%%%%%%%%%%%%%%%%%%

\bibliographystyle{utcaps}
\bibliography{biblio}

\end{document}